\documentclass[journal,article,submit,pdftex,moreauthors]{Definitions/mdpi}

\firstpage{1}
\pubvolume{1}
\issuenum{1}
\articlenumber{0}
\pubyear{2026}
\copyrightyear{2026}
\datereceived{}
\daterevised{}
\dateaccepted{}
\datepublished{}

\usepackage[english]{babel}
\usepackage{graphicx}
\usepackage{bm}
\usepackage{dsfont}
\usepackage{slashed}
\usepackage{xcolor}
\usepackage{subcaption}
\usepackage{comment}
\usepackage{float}
\usepackage[normalem]{ulem}

\newcommand{\itemuno}[1]{{\color{black} #1}}
\newcommand{\itemdos}[1]{{\color{black} #1}}
\newcommand{\itemtres}[1]{{\color{black} #1}}
\newcommand{\itemcuatro}[1]{{\color{black} #1}}
\newcommand{\itemcinco}[1]{{\color{black} #1}}

\newcommand{\agt}{\gtrsim}

\Title{Finite-Size Effects on the Critical End Point of Magnetized Quark Matter in the Nonlocal PNJL Model}

\Author{G. Lugones $^{1,*}$, S.A. Ferraris $^{2}$ and  A.G. Grunfeld $^{3,4}$}
\AuthorNames{G. Lugones, S.A. Ferraris and A.G. Grunfeld}

\address{%
$^{1}$ \quad Universidade Federal do ABC, Centro de Ci\^encias Naturais e Humanas, Avenida dos Estados 5001- Bang\'u, CEP 09210-580, Santo Andr\'e, SP, Brazil. \\
$^{2}$ \quad Physics Department, Comisión Nacional de Energía Atómica, Avenida del Libertador 8250, (1429) Buenos Aires, Argentina; \\
$^{3}$ \quad CONICET, Godoy Cruz 2290, (C1425FQB) Ciudad Autónoma de Buenos Aires, Argentina; \\
$^{4}$ \quad Instituto de Astronomía y Física del Espacio (IAFE, CONICET-UBA), Ciudad Universitaria C1428EGA, Ciudad Autónoma de Buenos Aires, Argentina; }

\corres{Correspondence: german.lugones@ufabc.edu.br}

\abstract{We investigate finite-size effects in the $T$--$\mu$ phase diagram
of magnetized quark matter within the framework of a nonlocal extension of
the Polyakov--Nambu--Jona-Lasinio (PNJL) model. Finite-size corrections are
incorporated through the multiple reflection expansion (MRE) formalism,
which describes a spherical quark droplet of radius $R$ and modifies the
density of states by including surface and curvature contributions. We
consider two-flavor quark matter at finite temperature and chemical
potential in the presence of a uniform magnetic field with strengths
ranging from $eB=0$ to $1$~GeV$^{2}$, and droplet radii from $R=3$~fm to
the bulk limit. The nonlocal PNJL (nlPNJL) model naturally reproduces both magnetic
catalysis at low temperatures and inverse magnetic catalysis near the
chiral transition, in agreement with lattice QCD results. We analyze the
chiral condensate, the traced Polyakov loop, the normalized quark
condensate, and the corresponding susceptibilities. 
\textcolor{black}{We find that finite-size effects do not modify the overall structure of the phase diagram, and that the coincidence of the chiral restoration and deconfinement transitions persists for all magnetic field strengths and system sizes explored, within the present implementation in which finite-size corrections are restricted to the fermionic sector.}
However, the critical end point (CEP) is notably shifted as a function of both the magnetic field strength and the system size: it moves toward higher chemical potentials and lower temperatures as the system size decreases, an effect that is significantly amplified by strong magnetic fields. Our results have potential implications for the physics of phase conversion in compact stars and for the interpretation of relativistic heavy-ion collision experiments.}

\begin{document}

\section{Introduction}

Strong external magnetic fields comparable to or exceeding the QCD confining
scale squared are expected to arise in several physical scenarios of current
interest, making the study of their impact on the QCD phase diagram a subject
of growing theoretical and phenomenological effort~\citep{Kharzeev:2012ph,Andersen:2014xxa,Miransky:2015ava}.
In compact astrophysical objects, the surface magnetic field of
magnetars is estimated at $\sim 10^{15}$~G~\citep{Duncan:1992hi}, 
and considerably higher values have been conjectured for their deep
interior, where some estimates suggest that the field strength could reach
$10^{18}$--$10^{19}$~G if dense quark matter is
present~\citep{Chatterjee:2014qsa}.
In relativistic heavy-ion collisions,
transient magnetic fields are generated by the ultrarelativistic motion of
the charged nuclei, reaching
$eB\sim 10^{18}$--$10^{19}$~G~$\approx (0.005$--$0.05)$~GeV$^{2}$ at RHIC
($\sqrt{s}\sim 200$~GeV per nucleon) and
$eB\sim 10^{19}$--$10^{20}$~G~$\approx (0.05$--$0.5)$~GeV$^{2}$ at the
LHC~\citep{Skokov:2009qp}. Although short-lived ($\sim 10^{-23}$~s), these
fields can affect the physics of the quark-gluon plasma, 
offering valuable opportunities to probe the phase diagram in the region of
deconfinement and chiral symmetry restoration. 

A common feature of the above scenarios is that the magnetized
quark matter under consideration is not an infinite bulk system. In the
interior of compact stars, a first-order deconfinement transition would
proceed through the nucleation and growth of quark matter droplets, whose
radii---ranging from a few to tens of femtometers---are set by the
competition between bulk, surface, and curvature energy
contributions~\citep{Lugones:2015bya}.
Likewise, the quark-gluon plasma produced in heavy-ion collisions has a
spatial extent of only a few femtometers, determined by the geometry of the
colliding nuclei. 
Finite-size effects can modify the behavior of
strongly interacting matter by restricting the available momentum space and
suppressing long-wavelength fluctuations, potentially shifting the location
of the CEP and altering the nature of the phase transition. Despite their
potential phenomenological relevance finite-size corrections have received
comparatively less attention than magnetic-field effects in the literature
on the QCD phase diagram. In this work, we address both effects
simultaneously by studying the $T$--$\mu$ phase diagram of magnetized quark
matter in a finite spherical geometry, within an effective model framework.

The nonperturbative character of strong interactions in this regime demands
the use of either lattice QCD (LQCD) techniques or effective theories capable
of capturing the essential features of QCD phenomenology. A prominent class
of effective approaches is based on the 
Nambu--Jona-Lasinio (NJL) model~\citep{Nambu:1961tp,Nambu:1961fr}, in which 
quarks interact through local four-point couplings that drive the spontaneous 
breakdown of chiral symmetry~\citep{Vogl:1991qt,Klevansky:1992qe,Hatsuda:1994pi}. 
A significant improvement over the local formulation is achieved by introducing nonlocal separable interactions, a feature supported by several effective approaches to QCD and known to yield results in closer agreement with LQCD. For a detailed overview of nonlocal NJL models and their applications to strongly interacting matter under extreme conditions, see Ref.~\citep{Dumm:2021vop}.
When supplemented with the coupling to a background temporal color gauge  field through the Polyakov loop $\Phi$~\citep{Fukushima:2003fw},  the resulting nlPNJL model~\citep{Contrera:2007wu,Hell:2008cc,Carlomagno:2013ona} provides a  simultaneous description of the chiral and deconfinement transitions whose pseudocritical temperatures are in good agreement with LQCD  results~\citep{Karsch:2003jg}.

A key feature of the QCD vacuum in the presence of an external magnetic field 
is the phenomenon of \emph{magnetic catalysis} (MC): at zero temperature, the 
chiral condensate grows with the field strength, an effect that is well 
established both in LQCD and in essentially all effective 
models. Near the chiral crossover, 
however, LQCD simulations with physical quark 
masses~\citep{Bali:2011qj,Bali:2012zg} reveal a qualitatively different 
behavior: the condensate becomes a non-monotonic function of $B$ and the 
pseudocritical temperature decreases with the field, an effect known as 
\emph{inverse magnetic catalysis} (IMC). Reproducing IMC has proven 
challenging for most local effective models, which generically predict an 
increase of the transition temperature with the 
field~\citep{Andersen:2014xxa,Kharzeev:2012ph,Miransky:2015ava}. Various 
remedies have been proposed, such as introducing explicit $B$-dependent 
couplings~\citep{Farias:2014eca, Ferreira:2014kpa}. In contrast, nlPNJL models 
have been shown to account for IMC already at the mean-field level, without 
any field-dependent modification of the model 
parameters~\citep{Pagura:2016pwr,dumm2017strong}. This natural emergence of 
IMC---rooted in the momentum dependence of the nonlocal quark 
self-energy---was demonstrated in the bulk limit at finite temperature $T$ and 
zero chemical potential $\mu$ in Ref.~\citep{dumm2017strong}, then at $T = 0$ and finite $\mu$ in Ref.~\citep{Ferraris:2021vun}, and subsequently 
extended to the full $T$--$\mu$ plane in 
Ref.~\citep{Carlomagno:2023clk}. 
These bulk studies showed in particular that the 
CEP temperature decreases monotonically with $eB$, in sharp contrast to the 
predictions of local NJL/PNJL 
models~\citep{Avancini:2012ee, Ferrari:2012yw, Costa:2013zca, Costa:2015bza}, 
and that the chiral and deconfinement transitions remain overlapped throughout the $T$--$\mu$--$B$ parameter space.

\itemcinco{In the past years, finite-volume effects in effective descriptions of
strongly interacting matter have been investigated using several
complementary approaches. Early studies within NJL-type models showed
that the presence of compact spatial dimensions or boundaries modifies
the pattern of chiral symmetry breaking and may affect the location and
even the order of the chiral transition
\cite{Abreu2006,Yasui2006,Abreu2011}. More recently, finite-volume
effects have been analyzed in PNJL and related Polyakov-loop extended
models, where the finite system size can affect both the chiral and
deconfinement sectors through the quark contribution to the
thermodynamic potential
\cite{Cristoforetti2010,Bhattacharyya:2012rp,Bhattacharyya:2014uxa,Pan2017}.}

\itemcinco{Different prescriptions have been employed to incorporate the finite
spatial extent of the system. These include Monte Carlo simulations
\cite{Cristoforetti2010}, functional-renormalization-group treatments
that account for the restriction of long-wavelength fluctuations
\cite{Tripolt2014,Braun2011}, compactification methods in which the
spatial momenta are discretized according to the boundary conditions
\cite{Abreu2011,Abreu2019}, and phenomenological implementations based
on a lower momentum cutoff in the thermodynamic potential
\cite{Bhattacharyya:2012rp,Pan2017}. In addition, finite-size effects
have been implemented through the Multiple Reflection Expansion (MRE) \cite{Balian:1970fw},
which modifies the density of states by surface and curvature
contributions and has been used to study thermodynamic quantities and
the chiral/deconfinement crossover in finite PNJL systems
\cite{Grunfeld:2017dfu}. Recent studies have further explored the
dependence of the phase diagram and the CEP on the geometry, boundary
conditions, and finite-volume prescription
\cite{MataCarrizal2022,CastanoYepes2022,Bernhardt2021}.
}

The present work builds on previous studies of the two-flavor nlPNJL model in the bulk, where strong-magnetic-field effects and the associated QCD phase structure were analyzed in Refs.~\citep{dumm2017strong, Ferraris:2021vun, Carlomagno:2023clk}. We extend those results by incorporating, for the first time within this framework, finite-size corrections through the MRE formalism. The MRE describes a spherical quark droplet of radius $R$ by modifying the density of states with surface and curvature contributions, thereby introducing an effective infrared cutoff that suppresses long-wavelength modes. This allows us to study the simultaneous interplay between magnetic-field effects, finite-size corrections, and the location of the CEP---a question that, to our knowledge, has not been addressed previously in the framework of nonlocal chiral quark models.

The article is organized as follows. In Sec.~\ref{sec:formalism} we describe
the theoretical formalism for magnetized quark matter within the nlPNJL
model, including the MRE treatment of finite-size effects. In
Sec.~\ref{sec:numsetup} we specify the model input and parameter choice, and
discuss the infrared structure of the MRE density of states. In
Sec.~\ref{sec:orderparams} we present our numerical results for the chiral
and deconfinement order parameters and susceptibilities, considering
different magnetic field strengths and system sizes. The resulting phase
diagram in the $T$--$\mu$ plane is discussed in
Sec.~\ref{sec:phase_diagram}. Finally, in Sec.~\ref{sec:conclusions} we
provide a summary and conclusions.

\section{PNJL Formalism with Finite-Size and Magnetic-Field Effects}
\label{sec:formalism}

In this work we consider the two-flavor nlPNJL model in the bulk, as developed in
Refs.~\citep{dumm2017strong,Ferraris:2021vun,Carlomagno:2023clk}, and extend it
by incorporating finite-size effects through the MRE. For completeness, we briefly review the bulk formalism to keep the
presentation self-contained and to establish the baseline against which
finite-size effects are assessed.

We consider quark matter composed of $u$ and $d$ quarks at finite temperature
$T$ and quark chemical potential $\mu$, in the presence of a uniform and static
external magnetic field $\bm{B}=B\,\hat z$. Confinement physics is modeled
through the coupling to a homogeneous temporal background color field, encoded
in the traced Polyakov loop $\Phi$.

For clarity, we organize the discussion as follows. We first summarize the bulk
mean-field thermodynamic potential in a magnetic background, then discuss its
regularization and the main thermodynamic observables, and finally introduce the
MRE treatment of finite-size effects for a spherical droplet.

\subsection{Bulk mean-field thermodynamic potential}

In the mean-field approximation (MFA), the regularized grand-canonical
potential is built from the bulk nlPNJL thermodynamic potential in the presence
of the magnetic field and the Polyakov-loop background; details of the
calculation can be found in
Refs.~\citep{dumm2017strong,Ferraris:2021vun,Carlomagno:2023clk}. In the
Polyakov gauge, one can take
\begin{equation}
\phi = \phi_3\,\lambda_3 + \phi_8\,\lambda_8 \, ,
\end{equation}
with $\phi_8=0$ for the cases considered here, so that
\begin{equation}
\phi_r=-\phi_g=\phi_3\, , \qquad \phi_b=0\, ,
\end{equation}
and the traced Polyakov loop becomes
\begin{equation}
\Phi = \frac{1}{3}\,\mathrm{Tr}\,e^{i\phi/T}
= \frac{1+2\cos(\phi_3/T)}{3}\, .
\end{equation}

The bulk MFA thermodynamic potential reads
\begin{align}
\Omega_{B,T,\mu}^{\rm MFA}
=&\; \frac{\bar\sigma^2}{2G}
- T \sum_{n=-\infty}^{\infty}\sum_{c=r,g,b}\sum_{f=u,d}
\frac{|q_f B|}{2\pi}
\int \frac{dp_3}{2\pi}
\Bigg[
\ln\!
\left( p_\parallel^2 + \big(M_{0,p_\parallel}^{\lambda_f,f}\big)^2 \right)
\nonumber \\
&\hspace{4.2cm}
+ \sum_{k=1}^{\infty} \ln \Delta_{k,p_\parallel}^{f}
\Bigg]
+ \mathcal U(\Phi,T)\, ,
\label{eq:OmegaMFA_bulk_rewrite}
\end{align}
where
\begin{align}
\Delta_{k,p_\parallel}^{f}
=&\; \left( 2k|q_f B| + p_\parallel^2
+ M_{k,p_\parallel}^{+,f} M_{k,p_\parallel}^{-,f} \right)^2
 + p_\parallel^2
\left( M_{k,p_\parallel}^{+,f} - M_{k,p_\parallel}^{-,f} \right)^2 .
\label{eq:Delta_bulk_rewrite}
\end{align}
The momentum-dependent constituent masses are given by
\begin{equation}
M_{k,p_\parallel}^{\lambda,f}
= (1-\delta_{k_{\lambda},-1})\,m_c
+ \bar\sigma\,g_{k,p_\parallel}^{\lambda,f}\, ,
\label{eq:Mkp_rewrite}
\end{equation}
with
\begin{align}
g_{k,p_\parallel}^{\lambda,f}
=&\; \frac{4\pi}{|q_f B|}(-1)^{k_\lambda}
\int \frac{d^2p_\perp}{(2\pi)^2}
\,g(p_\perp^2+p_\parallel^2)
\,e^{-p_\perp^2/B_f}
L_{k_\lambda}\!\left(\frac{2p_\perp^2}{B_f}\right) .
\label{eq:gkp_rewrite}
\end{align}
Here $L_n(x)$ denotes the Laguerre polynomial, $g(p^2)$ is the Fourier
transform of the nonlocal form factor ${\cal G}(z)$ introduced in the model
action, and we use the definitions
\begin{equation}
k_{\pm}=k-\frac{1}{2}\pm\frac{s_f}{2}\, ,
\qquad
s_f=\mathrm{sign}(q_f B)\, ,
\qquad
B_f=|q_f B|\, .
\end{equation}
Moreover,
\begin{equation}
p_\parallel \equiv \big(p_3,\,\omega_n-i\mu+\phi_c\big)
\qquad \Rightarrow \qquad
p_\parallel^2=p_3^2+\big(\omega_n-i\mu+\phi_c\big)^2 ,
\end{equation}
where $\omega_n=(2n+1)\pi T$ are the fermionic Matsubara frequencies. The
external magnetic field $\bm{B}$ is taken to point along the 3-axis.

The Polyakov-loop effective potential is taken in the polynomial form of
Refs.~\cite{Ratti:2005jh,Schaefer:2007pw},
\begin{equation}
\frac{\mathcal U(\Phi,T)}{T^4}
= -\frac{b_2(T)}{2}\,\Phi^2
-\frac{b_3}{3}\,\Phi^3
+\frac{b_4}{4}\,\Phi^4\, ,
\label{eq:polyakov_potential_rewrite}
\end{equation}
with
\begin{equation}
b_2(T)=a_0+a_1\left(\frac{T_0}{T}\right)
+a_2\left(\frac{T_0}{T}\right)^2
+a_3\left(\frac{T_0}{T}\right)^3 .
\end{equation}
As in Ref.~\cite{Carlomagno:2023clk}, we use
$a_0=6.75$, $a_1=-1.95$, $a_2=2.625$, $a_3=-7.44$, $b_3=0.75$,
$b_4=7.5$, and $T_0=210\,$MeV.

\subsection{Regularization and bulk thermodynamic observables}

The momentum integral in Eq.~\eqref{eq:OmegaMFA_bulk_rewrite} is ultraviolet
divergent. We regularize it by subtracting the corresponding free contribution
and adding it back in a regularized form, namely
\begin{equation}
\Omega_{B,T,\mu}^{\rm MFA,reg}
= \Omega_{B,T,\mu}^{\rm MFA}
- \Omega_{B,T,\mu}^{\rm free}
+ \Omega_{B,T,\mu}^{\rm free,reg} .
\label{eq:Omega_reg_rewrite}
\end{equation}
Here $\Omega_{B,T,\mu}^{\rm free}$ is obtained from
Eq.~\eqref{eq:OmegaMFA_bulk_rewrite} by setting $\bar\sigma=0$, while keeping
the coupling to both the magnetic field and the Polyakov-loop background. The
regularized free term can be written as
\begin{equation}
\Omega_{B,T,\mu}^{\rm free,reg}
= \Omega_{\rm mag} + \Omega_{\rm free\,gas} ,
\end{equation}
where~\cite{Menezes:2008qt}
\begin{equation}
\Omega_{\rm mag}
= -\frac{3}{2\pi^2}\sum_f (q_f B)^2
\left[\zeta'(-1,x_f)+F(x_f)\right] ,
\label{eq:Omega_mag_rewrite}
\end{equation}
\begin{equation}
\Omega_{\rm free\,gas}
= -T\sum_{c,f,k}\frac{|q_f B|}{2\pi}\,\alpha_k
\int \frac{dp}{2\pi}
\,G_{k,p}^{f}(\phi_c,\mu,T) ,
\label{eq:Omega_free_gas_rewrite}
\end{equation}
with
\begin{align}
F(x_f)
&= \frac{x_f^2}{4}
-\frac{1}{2}(x_f^2-x_f)\ln x_f , \\
G_{k,p}^{f}(\phi_c,\mu,T)
&= \sum_{s=\pm}
\ln\!\left[1+
\exp\!\left(-\frac{\epsilon_{kp}^{f}+i\phi_c+s\mu}{T}\right)
\right] .
\end{align}
We have also defined
\begin{equation}
x_f = \frac{m_c^2}{2|q_f B|}\, ,
\qquad
\alpha_k = 2-\delta_{k0}\, ,
\qquad
\epsilon_{kp}^{f} = \sqrt{p^2+2k|q_f B|+m_c^2}\, .
\end{equation}

The mean fields $\bar\sigma(B,T,\mu)$ and $\Phi(B,T,\mu)$ are obtained from
the stationarity conditions
\begin{equation}
\frac{\partial \Omega_{B,T,\mu}^{\rm MFA,reg}}{\partial \bar\sigma}=0,
\qquad
\frac{\partial \Omega_{B,T,\mu}^{\rm MFA,reg}}{\partial \Phi}=0.
\label{eq:gap_rewrite}
\end{equation}
Once the solution of these gap equations is found, other bulk observables
follow in the standard way. In particular, the regularized quark condensate of
flavor $f$ is
\begin{equation}
\langle \bar q_f q_f \rangle_{B,T,\mu}^{\rm reg}
= \frac{\partial \Omega_{B,T,\mu}^{\rm MFA,reg}}{\partial m_c} .
\label{eq:condensate_rewrite}
\end{equation}

Here we use the normalized flavor-averaged condensate, as in
Ref.~\citep{dumm2017strong},
\begin{equation}
\bar\Sigma_{B,T}
= \frac{1}{2}\left(\Sigma^u_{B,T}+\Sigma^d_{B,T}\right) ,
\end{equation}
where
\begin{equation}
\Sigma^f_{B,T}
= -\frac{2m_c}{S^4}
\left[
\langle \bar q_f q_f \rangle_{B,T,\mu}^{\rm reg}
-\langle \bar q_f q_f \rangle_{0,0,0}^{\rm reg}
\right] ,
\end{equation}
and $S=(135\times 86)^{1/2}\,$MeV is a phenomenological scale.

The pseudocritical chiral and deconfinement temperatures are located from the
maxima of the corresponding susceptibilities,
\begin{equation}
\chi_{\rm ch}
= -\frac{\partial}{\partial T}
\left[
\frac{\langle \bar u u\rangle_{B,T,\mu}^{\rm reg}
+\langle \bar d d\rangle_{B,T,\mu}^{\rm reg}}{2}
\right] ,
\qquad
\chi_{\Phi}=\frac{\partial \Phi}{\partial T} .
\label{susceptibilities}
\end{equation}
With this definition, purely magnetic vacuum terms that do not depend on $T$
do not contribute to $\chi_{\rm ch}$.

In the crossover regime, the location of the susceptibility peaks defines a
\emph{pseudocritical temperature} $T_{\rm pc}$. In the first-order regime, the
transition temperature $T_{\rm tr}$ is instead identified by the degeneracy of
the two minima of the thermodynamic potential. For brevity, where the
distinction is not essential we use the generic term ``transition temperature''
to refer to both cases.

\subsection{Finite-size effects: MRE description of a spherical droplet}

To incorporate finite-size effects we employ the MRE formalism for a spherical
droplet of radius $R$~\cite{Grunfeld:2017dfu, Grunfeld:2024ihq}.
In this approach the density of states is modified according to
\begin{equation}
\rho_{\mathrm{MRE},f}(p)
= 1 + \frac{6\pi^2}{pR}\,f_{S,f}(p)
+ \frac{12\pi^2}{(pR)^2}\,f_{C,f}(p) .
\label{eq:rhoMRE_rewrite}
\end{equation}
The surface contribution is
\begin{equation}
f_{S,f}(p)
= -\frac{1}{8\pi}
\left(1-\frac{2}{\pi}\arctan\frac{p}{m_c}\right) ,
\label{eq:fS_rewrite}
\end{equation}
and for the curvature term we use the standard Madsen ansatz\footnote{The Madsen ansatz~\cite{Madsen:1994vp} for the curvature term smoothly interpolates between the massless MIT-bag limit, $f_C=-1/(24\pi^2)$, and the nonrelativistic Dirichlet limit, $f_C=1/(12\pi^2)$, and was shown to reproduce shell-model calculations for massive quarks. A different choice of boundary condition would modify the numerical values of $f_S$ and $f_C$ but not the bulk Landau-level structure or the infrared mechanism discussed below, so the qualitative trends reported in this work are expected to be robust.},
\begin{equation}
f_{C,f}(p)
= \frac{1}{12\pi^2}
\left[1-\frac{3p}{2m_c}
\left(\frac{\pi}{2}-\arctan\frac{p}{m_c}\right)
\right] .
\label{eq:fC_rewrite}
\end{equation}
As emphasized in the MRE literature, $\rho_{\mathrm{MRE},f}(p)$ can become
negative at sufficiently small momenta because higher-order terms in $1/R$ are
neglected. To avoid this unphysical region, an infrared cutoff is introduced by
requiring
\begin{equation}
\rho_{\mathrm{MRE},f}(p)=0 ,
\end{equation}
and taking $\Lambda_{\rm IR}$ as the largest root of this equation.

\itemdos{
A few comments on the interpretation of this cutoff are in order. The
suppression of long-wavelength modes is a physical finite-size effect: modes
with wavelengths comparable to, or larger than, the droplet size do not fit
inside the system and are therefore strongly
affected by the boundary. The limitation of the MRE lies not in this
infrared suppression itself, but in the fact that the MRE is a truncated
semiclassical expansion in powers of \(1/(pR)\). It is therefore quantitatively
controlled for \(pR\gg 1\), while for \(p\lesssim 1/R\) higher-order terms become
important. The negative density of states that appears at small momenta is a
manifestation of this loss of control, and the infrared cutoff is a standard
prescription used to exclude this unphysical region.}

In the absence of a magnetic field, the replacement in a generic bulk momentum
integral is
\begin{equation}
\frac{1}{(2\pi)^3}\int d^3p\;\cdots
\;
\longrightarrow
\;
\frac{1}{(2\pi)^3}
\int_{\Lambda_{\rm IR}}^{\infty} 4\pi p^2\,dp\;
\rho_{\mathrm{MRE},f}(p)\;\cdots .
\label{eq:MRE_replacement_rewrite}
\end{equation}
In the presence of the magnetic field, the transverse motion is quantized into
Landau levels and
\begin{equation}
p = \sqrt{p_3^2 + 2k|q_f B|}\, .
\label{eq:ecuacion29}
\end{equation}
Hence the relevant substitution in thermodynamic integrals becomes\footnote{It is worth stressing that the use of the MRE density of states in a magnetic field is a semiclassical prescription. The magnetic field is included through the Landau-level kinetic momentum \eqref{eq:ecuacion29} and the corresponding degeneracy factor, while the standard surface and curvature MRE functions are retained. A fully microscopic derivation of the spectral density in a finite magnetized domain could in principle generate additional terms depending on the magnetic length \(\ell_B=1/\sqrt{|q_fB|}\) and on the orientation of the boundary with respect to \(\mathbf B\). A systematic derivation and analysis of such anisotropic boundary corrections is beyond the scope and objectives of the present article.}
\begin{equation}
\frac{|q_f B|}{2\pi}
\sum_{k=0}^{\infty}\alpha_k
\int\frac{dp_3}{2\pi}\;\cdots
\;
\longrightarrow
\;
\frac{|q_f B|}{2\pi}
\sum_{k=0}^{\infty}\alpha_k
\int_{\Lambda_{\rm IR}}^{\infty}\frac{dp_3}{2\pi}
\,\rho_{\mathrm{MRE},f}(p)\;\cdots ,
\label{eq:MRE_B_rewrite}
\end{equation}
where $\Lambda_{\rm IR}$ is the largest solution of
$\rho_{\mathrm{MRE},f}(\sqrt{p_3^2+2k|q_f B|})=0$ with respect to $p_3$.

\itemuno{
Within the present implementation, the MRE correction is applied only to the
fermionic sector: the Polyakov-loop potential $\mathcal U(\Phi,T)$ is left
unchanged and treated as a bulk gluonic effective potential. This choice is not
merely a simplifying assumption; it follows the standard practice adopted in
PNJL studies of finite-volume effects, where the Polyakov-loop potential is
calibrated to pure-gauge lattice thermodynamics in the bulk and kept in that
form when finite-size corrections are introduced in the fermionic sector
through low-momentum cutoffs, MRE, or related
prescriptions~\citep{Bhattacharyya:2012rp, Bhattacharyya:2014uxa, Grunfeld:2017dfu}.
This procedure is supported on physical grounds by two complementary arguments.

First, as shown by Sasaki and Redlich~\citep{Sasaki:2012bi}, the effective
Polyakov-loop potential in SU(3) Yang-Mills theory can be derived from the
partition function using the background field method, where the $n$-body gluon
contributions are obtained by performing the integration over the color gauge
group. The resulting potential is entirely determined by the group-theoretical
structure of the Haar measure. Because $\mathcal U(\Phi,T)$ originates from
this algebraic color integration rather than from a momentum-space integral
over propagating modes, it is not subject to the modification of the density
of states that the MRE formalism introduces: the surface and curvature
corrections in Eq.~(\ref{eq:rhoMRE_rewrite}) have no counterpart in the
group-integration procedure that generates $\mathcal U(\Phi,T)$.

Second, on dimensional grounds, the gluonic sector is governed by the
confinement scale $\Lambda_{\rm QCD}\sim 200$--$300$~MeV, which corresponds to
a correlation length $\xi_g\lesssim 1$~fm, well below the smallest droplet
radius considered here ($R=3$~fm). In contrast, quarks are affected more
directly by the finite size of the system because their allowed momenta depend
on the spatial extension of the droplet. It is therefore
physically reasonable to expect that the leading finite-size corrections enter
through the quark sector. We stress, however, that this is a plausibility
argument rather than a derivation: a fully consistent treatment of finite-volume
effects in the gluonic sector---which would require, for instance, a
determination of the $R$-dependence of the parameters $T_0$, $a_i$, and $b_i$
entering Eq.~(\ref{eq:polyakov_potential_rewrite})---lies beyond the scope of
bulk-fitted PNJL-type effective potentials.

It is important to be explicit about the implications of this treatment for
the interpretation of our results. Because $\mathcal U(\Phi,T)$ is calibrated
in the bulk limit, it is not sensitive to the geometric suppression of
infrared modes that the MRE imposes on the quark sector, and the
$R$-dependence of $\Phi$ in our calculation arises indirectly through its
coupling to the quark determinant. As a consequence, the coincidence of the
chiral and deconfinement pseudocritical temperatures observed throughout this
work---a result well established in the bulk
limit~\citep{Carlomagno:2023clk}---should not be regarded as an independent
prediction of the finite-volume dynamics of the gluonic sector, but rather as
a feature of the present framework that survives the introduction of the
fermionic finite-size corrections. A genuine assessment of whether this
coincidence persists in a fully finite-volume treatment of the pure-gauge
sector would require a finite-size formulation of $\mathcal U(\Phi,T)$ itself,
which is currently not available.
}

Having established the MRE replacement rules and the treatment of the
Polyakov-loop potential, we can now write the finite-size corrected
thermodynamic potential in a compact form. For this purpose it is convenient
to define the logarithmic kernel
\begin{align}
\mathcal F_{k,p_\parallel}^{f}
=&\;
\ln\!\left[
\frac{p_\parallel^2+m_c^2}
{p_\parallel^2+\big(M_{0,p_\parallel}^{\lambda_f,f}\big)^2}
\right]
+ \sum_{k=1}^{\infty}
\ln\!\left[
\frac{\big(2k|q_f B|+p_\parallel^2+m_c^2\big)^2}
{\Delta_{k,p_\parallel}^{f}}
\right] ,
\label{eq:Fkernel_rewrite}
\end{align}
which encodes the difference between the interacting and free quark
contributions for each Landau level $k$ and longitudinal momentum $p_\parallel$,
as required by the regularization procedure of
Eq.~\eqref{eq:Omega_reg_rewrite}. With this definition, the regularized
finite-size thermodynamic potential takes the form
\begin{align}
\Omega_{\rm reg}^{\rm MRE}
=&\; \frac{\bar\sigma^2}{2G}
+ \mathcal U(\Phi,T)
+ \Omega_{\rm mag}
+ T\sum_{n,c,f}\frac{|q_fB|}{\pi}
\int_{\Lambda_{\rm IR}}^{\infty}\frac{dp_3}{2\pi}
\,\rho_{\mathrm{MRE},f}(p)
\,\mathcal F_{k,p_\parallel}^{f}
\nonumber \\
&\;
- T\sum_{c,f}\frac{|q_fB|}{2\pi}
\sum_{k=0}^{\infty}\alpha_k
\int_{\Lambda_{\rm IR}}^{\infty}\frac{dp_3}{2\pi}
\,\rho_{\mathrm{MRE},f}(p)
\,G_{k,p}^{f}(\phi_c,\mu,T) .
\label{eq:OmegaMRE_schematic_rewrite}
\end{align}
The first line contains the mean-field condensate energy, the Polyakov-loop
potential, the magnetic vacuum contribution, and the MRE-corrected
regularization of the quark determinant; the second line is the finite-size
corrected free-gas contribution. 

Finally, we remark that the model parameters entering the nonlocal form factor
and the coupling constants are fixed in bulk vacuum,
\begin{equation}
T=0\, , \qquad \mu=0\, , \qquad eB=0\, ,
\end{equation}
and then kept unchanged throughout the finite-$T$, finite-$\mu$, finite-$B$,
and finite-size analysis.

\section{Model input and numerical setup}
\label{sec:numsetup}

In this section we present the numerical results obtained from the formalism introduced above. Before discussing the behavior of the order parameters and the corresponding phase structure, it is necessary to specify the model input, namely the nonlocal form factor and the parameter sets used in the calculations.

\subsection{Model input and parameter choice}
\label{subsec:parameters}

In order to obtain definite numerical predictions, one has to specify the functional form of the nonlocal form factor \(g(p^2)\) appearing in Eq.~(\ref{eq:gkp_rewrite}), together with the model parameters \(m_c\), \(G\), and \(\Lambda\). In what follows we adopt the Gaussian ansatz
\begin{equation}
g(p^{2})=\exp\!\left(-p^{2}/\Lambda^{2}\right)\, .
\end{equation}
This choice is widely used in nonlocal chiral quark models and has the practical advantage that the integral entering Eq.~\eqref{eq:gkp_rewrite} can be carried out analytically, which considerably simplifies the subsequent numerical calculations. In addition, previous analyses within this class of models indicate that the qualitative features of the results are not expected to depend strongly on the specific shape of the form factor.  

For the Gaussian form factor, the effective constituent quark masses introduced in Eq.~(\ref{eq:Mkp_rewrite}) take the form~\citep{dumm2017strong}
\begin{align}
M_{k,p_\parallel}^{\pm,f}
&=
(1-\delta_{k_\pm,-1})\,m_c
 +
\bar{\sigma}\,
\frac{\left(1-|q_f B|/\Lambda^2\right)^{k_\pm}}
     {\left(1+|q_f B|/\Lambda^2\right)^{k_\pm+1}}
\exp\!\left(-p_\parallel^2/\Lambda^2\right) .
\label{eq:mass_gaussian_num}
\end{align}
Here \(\Lambda\) sets the effective soft momentum scale associated with the nonlocal interaction and therefore plays the role of an ultraviolet regulator in the model.

We fix the model parameters so as to reproduce physical observables at $\mu = B = 0$, 
adopting $-\langle \bar{q} q \rangle^{1/3} = 230\,\text{MeV}$. 
This leads to $m_c = 6.4\,\text{MeV}$, $\Lambda = 677.8\,\text{MeV}$, and $G\Lambda^2 = 23.65$~\cite{Ferraris:2021vun}, 
yielding $m_\pi = 139\,\text{MeV}$ and $f_\pi = 92.4\,\text{MeV}$.

Once a parameter set has been selected, the coupled gap equations, Eq.(\ref{eq:gap_rewrite}), can be solved numerically for given values of the temperature \(T\), chemical potential \(\mu\), and magnetic field \(B\). As expected, for some regions of the \((T,\mu)\) plane and for fixed magnetic field, more than one solution may coexist. In those cases, the physically realized state is identified with the solution that minimizes the corresponding thermodynamic potential.

The results presented below have been obtained following this prescription.

\begin{table}[tb]
\centering
\caption{Representative values of the infrared cutoff $\Lambda_{\rm IR}$ for light quarks as a function of the droplet radius $R$, using a common current mass $m_c = 6.5~\mathrm{MeV}$ for the $u$ and $d$ flavors. In the bulk limit, $R\to\infty$, one recovers $\Lambda_{\rm IR}=0$.}
\label{tab:cutoff_rewrite}
\begin{tabular}{cc}
\hline\hline
$R$ [fm] & $\Lambda_{\rm IR}$ [MeV] \\
\hline
3        & 52.53 \\
5        & 33.54 \\
10       & 18.88 \\
$\infty$ & 0     \\
\hline\hline
\end{tabular}
\end{table}

\subsection{Infrared cutoff in the MRE density of states}
\label{subsec:infrared_mre}

\begin{figure}[tb]
\centering
\includegraphics[width=0.55\textwidth]{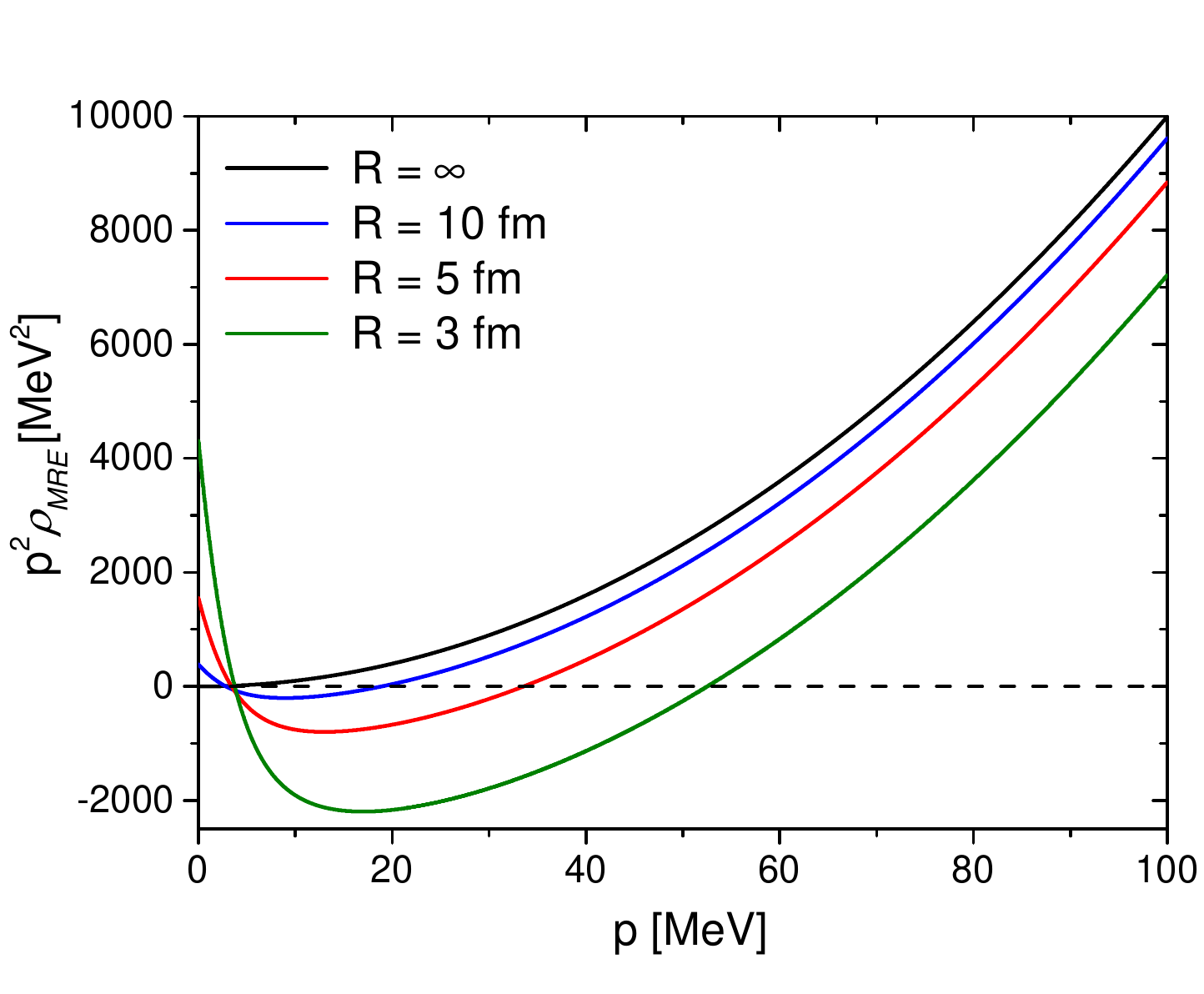}
\caption{Behavior of \(p^{2}\rho_{\rm MRE}(p)\) as a function of the momentum
for the LLL and several droplet radii. }
\label{figuracutoff}
\end{figure}

Before presenting the thermodynamic results, it is worth discussing the
infrared behavior of the MRE density of states in the presence of a magnetic
field, since this feature is relevant for the numerical implementation of the
formalism. In Fig.~\ref{figuracutoff} we show the quantity
\(p^{2}\rho_{\rm MRE}(p)\) for the lowest Landau level (LLL, \(k=0\)) and for
several droplet radii, \(R=3\), \(5\), \(10\)~fm and the bulk limit.  As discussed below, the infrared cutoff $\Lambda_{\rm IR}$ arises exclusively in the LLL; since the MRE density of states for $k=0$ does not depend on the
magnetic field strength, the values of $\Lambda_{\rm IR}$ listed in
Table~\ref{tab:cutoff_rewrite} are determined solely by the droplet radius
and the current quark mass, and remain unchanged for all values of $eB$
considered in this work.

For finite-size droplets, the function \(p^{2}\rho_{\rm MRE}(p)\) develops a
zero in the low-momentum region and becomes negative below it. This is the
well-known infrared pathology of the MRE approach, associated with the fact
that the semiclassical density of states ceases to be reliable when surface
corrections become too large compared with the bulk contribution. In order to
exclude this unphysical region, one introduces an infrared cutoff
\(\Lambda_{\rm IR}\), defined as the largest root of
\begin{equation}
p^{2}\rho_{\rm MRE}(p)=0 \, .
\end{equation}
The momentum integrals are then restricted to the domain
\(p>\Lambda_{\rm IR}\), where the density of states remains positive.

As expected, the value of \(\Lambda_{\rm IR}\) depends strongly on the droplet
radius. Smaller systems lead to larger cutoff values, reflecting a stronger
suppression of long-wavelength modes by finite-size effects. For the
representative cases considered here, one finds
\(\Lambda_{\rm IR}\simeq 52.5\), \(33.5\), and \(18.9\) MeV for
\(R=3\), \(5\), and \(10\) fm, respectively, while the cutoff moves close to
the origin for \(R=100\) fm and vanishes in the bulk limit.

A distinctive feature of the present problem is that this infrared zero
appears only in the LLL. For higher Landau levels
(\(k\geq 1\)), the corresponding function \(p^{2}\rho_{\rm MRE}(p)\) remains
positive over the whole momentum range for all radii analyzed here, so that
no infrared cutoff is required and one has \(\Lambda_{\rm IR}=0\). This
special role of the LLL can be understood as a consequence of the effective
dimensional reduction induced by the magnetic field: since the low-energy
dynamics in the LLL is essentially restricted to the direction parallel to the
field, boundary effects become comparatively more important, which enhances
the sensitivity of this sector to finite-size corrections.

Therefore, the infrared cutoff introduced by the MRE formalism is not a
generic feature of all Landau levels, but a specific finite-size effect of the
LLL sector. This point is important for the discussion below, since it shows
that the modification of the available phase space is highly selective and
acts predominantly on the sector that dominates the low-energy dynamics in strong
magnetic fields.

\section{Order parameters and susceptibilities}
\label{sec:orderparams}

In this section we present the numerical results for the chiral and
deconfinement order parameters as functions of temperature, for different
values of the droplet radius $R$, the magnetic field strength $eB$, and
the quark chemical potential $\mu$. All figures in this section share a
common panel layout: the left column displays finite-size effects at
$eB=0$ for several droplet radii ($R=\infty$, $10$, $5$, and $3$~fm),
while the right column shows the magnetic-field dependence in the bulk
limit ($eB=0$, $0.1$, $0.5$, and $1.0$~GeV$^{2}$). Rows correspond to
$\mu=0$, $75$, and $150$~MeV (except for the susceptibilities, where
only $\mu=0$ and $75$~MeV are shown). The bulk nlPNJL framework and some
aspects of its thermodynamics were introduced in
Refs.~\cite{dumm2017strong} and ~\cite{Carlomagno:2023clk}, where different parameter sets were
considered. Here we restrict ourselves to a single representative
parametrization but carry out a more detailed analysis of the bulk sector,
presenting a systematic study of the chiral condensate, the Polyakov loop,
the normalized quark condensate, and the corresponding susceptibilities as
functions of temperature, chemical potential, and magnetic field strength.
This extended bulk characterization serves both as a complement to the
results of Refs.~\cite{dumm2017strong} and ~\cite{Carlomagno:2023clk} and as the baseline against
which finite-size effects are identified and quantified throughout this
work.

\begin{figure}[tb]
\centering
\includegraphics[width=0.80\textwidth,trim={0cm 1.20cm 0cm 1.0cm}]{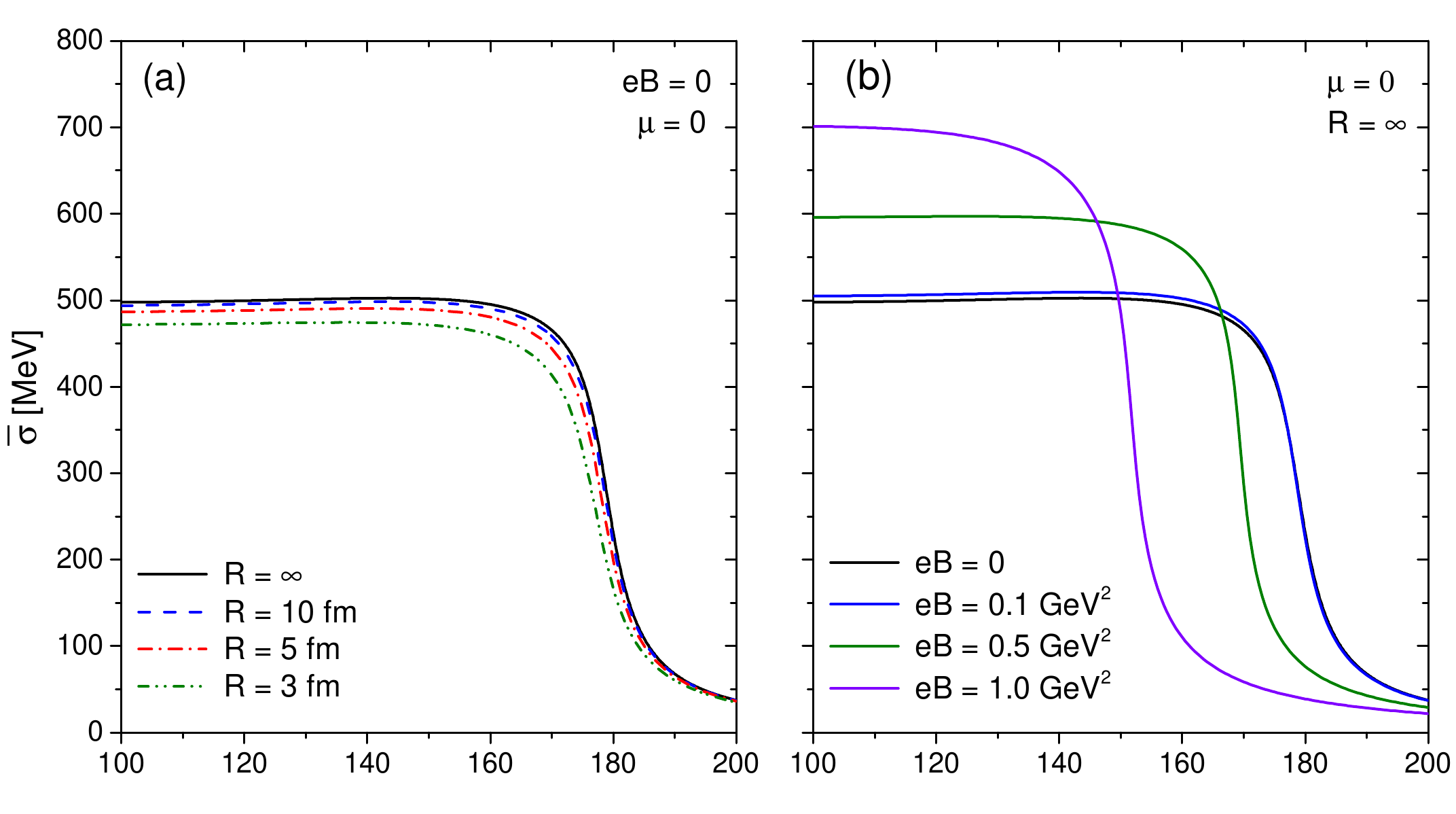} 
\includegraphics[width=0.80\textwidth,trim={0cm 1.20cm 0cm 0.0cm}]{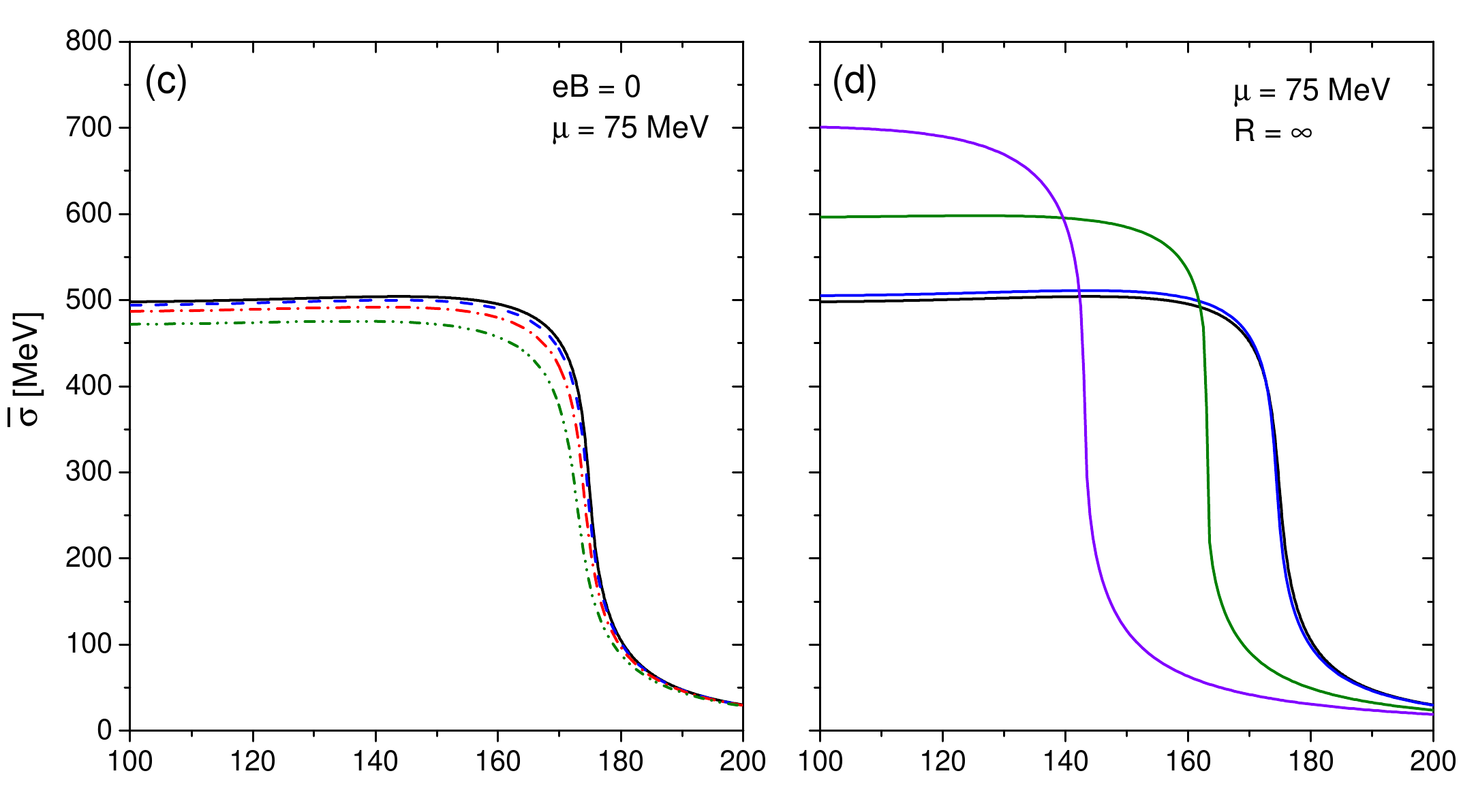} 
\includegraphics[width=0.80\textwidth,trim={0cm 0.00cm 0cm 0.0cm}]{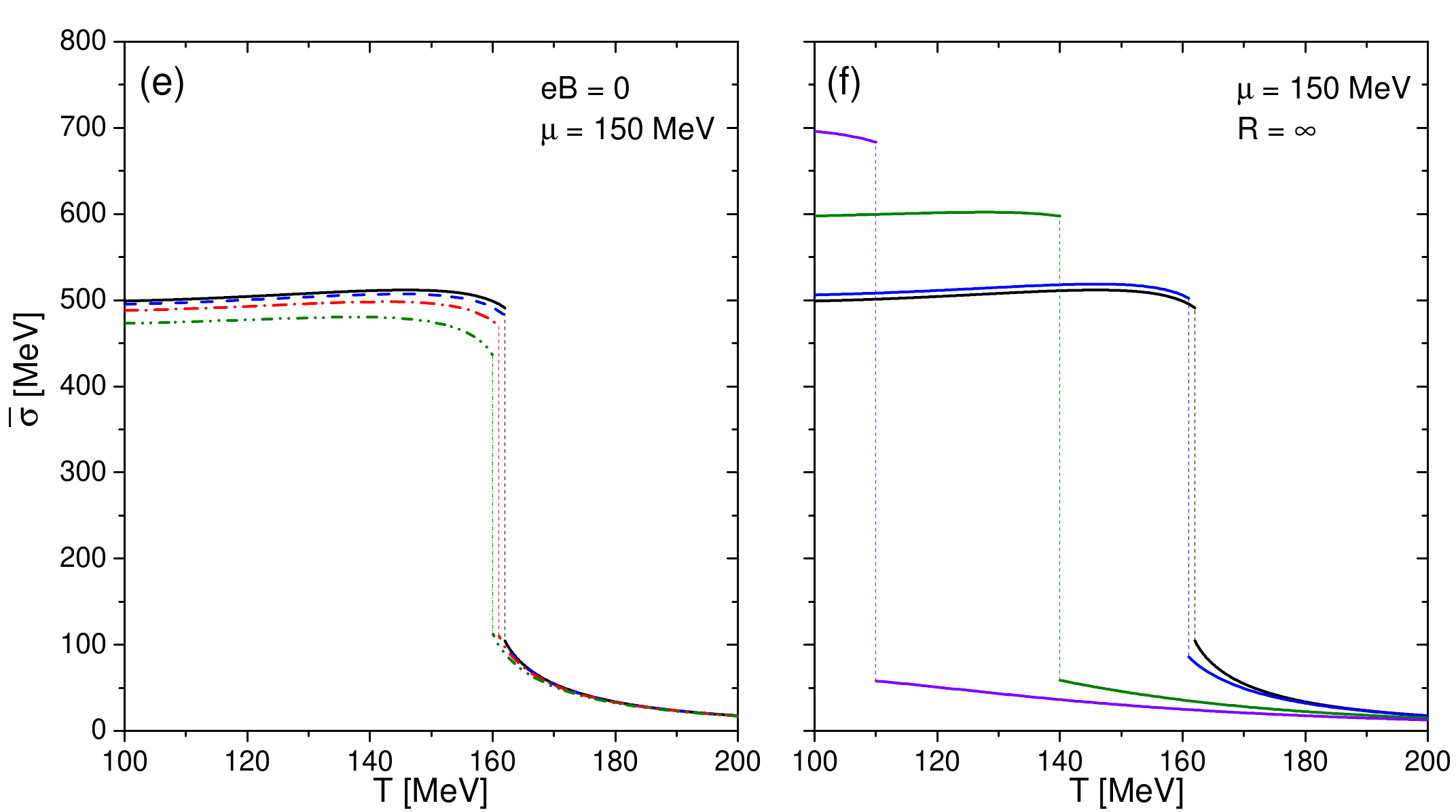} 
\caption{Mean-field $\bar\sigma$ as a function of the temperature $T$.
Left column: finite-size effects at $eB=0$ for several droplet radii.
Right column: bulk limit ($R=\infty$) for different magnetic field strengths.
Rows correspond to $\mu=0$ [panels~(a) and~(b)], $\mu=75$~MeV
[panels~(c) and~(d)], and $\mu=150$~MeV [panels~(e) and~(f)].
The left column illustrates the systematic shift of the crossover to lower
temperatures with decreasing system size, while the right column displays the
interplay between magnetic catalysis at low $T$ and inverse magnetic catalysis
near the pseudocritical temperature. At $\mu=150$~MeV, the transition
enters the first-order regime.}
\label{fig:sigma_panels}
\end{figure}

\subsection{Mean-field $\bar\sigma$}
\label{subsec:sigma_results}

We begin by analyzing the behavior of the mean-field
$\bar\sigma$---the solution of the gap equation~(\ref{eq:gap_rewrite})---as
a function of temperature for different values of $R$, $eB$, and $\mu$.
The results are summarized in the six panels of
Fig.~\ref{fig:sigma_panels}.

Three distinct mechanisms are found to modify the chiral transition:
finite-size effects, the external magnetic field, and the quark chemical
potential. We discuss each in turn, drawing on the figure panels for
illustration.

Finite-size effects lower the chiral condensate at all temperatures and shift the crossover to lower temperatures, anticipating the partial restoration of chiral symmetry. At $\mu=0$ and $eB=0$ [panel~(a)], the effect is modest for $R=10$~fm but becomes clearly visible for $R=3$~fm, where the vacuum
value of $\bar\sigma$ is reduced by approximately $5\%$ and the crossover
shifts to noticeably lower $T$. Despite this shift, the transition remains a
smooth crossover for all radii---no change in the order of the transition is
induced purely by geometric confinement finite size effects. The same pattern persists at
$\mu=75$~MeV [panel~(c)], where the crossover is steeper but the
hierarchical ordering with $R$ is preserved. At $\mu=150$~MeV [panel~(e)], the transition is already first order in the bulk and remains so for all radii; as in the previous cases, reducing $R$ anticipates the partial restoration of chiral symmetry by shifting the transition to lower temperatures.

The magnetic field plays a dual role. At low temperatures it enhances
$\bar\sigma$ through \emph{magnetic catalysis}: the quantization of
transverse motion into Landau levels increases the degeneracy of low-energy
states---particularly in the LLL, whose density of states
grows linearly with $|q_f B|$---thereby strengthening the infrared dynamics
that drives chiral symmetry breaking and leading to a larger value of the
mean field $\bar\sigma$. This is clearly seen in the right column of
Fig.~\ref{fig:sigma_panels}, where the low-$T$ value of $\bar\sigma$ rises
from $\sim 500$~MeV at $eB=0$ to $\sim 700$~MeV at $eB=1.0$~GeV$^{2}$.
Near the transition, however, the behavior reverses: the pseudocritical
temperature $T_{\rm pc}$ decreases monotonically with increasing $eB$---from
$\sim 180$~MeV at $eB=0$ to $\sim 145$~MeV at $eB=1.0$~GeV$^{2}$ for
$\mu=0$ [panel~(b)]---constituting the 
IMC effect. The competition between MC and IMC produces a characteristic
crossing of the curves: for strong fields, $\bar\sigma(T)$ starts well above
the $eB=0$ curve but drops below it near $T_{\rm pc}$
[panels~(b) and~(d)]. It is remarkable that the nonlocal PNJL framework
reproduces this behavior
naturally~\cite{Pagura:2016pwr,dumm2017strong}, in agreement with
lattice-QCD observations~\cite{Bali:2011qj,Bali:2012zg}, without ad hoc
modifications to the coupling constant or the Polyakov-loop potential.
For moderate fields ($eB=0.1$~GeV$^{2}$), the curves remain very close to
the $eB=0$ result, indicating that magnetic effects become phenomenologically
significant only for $eB\agt 0.1$~GeV$^{2}$.

A finite chemical potential steepens the crossover and lowers the transition
temperature, pushing the system toward the first-order regime. At
$\mu=150$~MeV and strong magnetic fields [panel~(f)], the transition is
unambiguously first-order, with $\bar\sigma$ exhibiting a clear
discontinuity and the low-temperature enhancement due to MC being even more
pronounced than at lower $\mu$.

These three effects act cumulatively: at a fixed point in the $T$--$\mu$
plane, reducing the system size, increasing the magnetic field, or raising
the chemical potential can each push the system from the crossover regime
into the first-order region, as the CEP shifts toward higher $\mu$ and
lower $T$. It is important to note, however, that finite-size effects
preserve the qualitative topology of the phase diagram: for all values of
$R$ and $eB$ explored in this work, the transition remains a crossover at
low chemical potentials and becomes first order at larger $\mu$, with a CEP
separating both regimes. What changes is the location of the CEP, not the
overall structure of the phase diagram.


\begin{figure}[tb]
\centering
\includegraphics[width=0.80\textwidth,trim={0cm 1.20cm 0cm 1.0cm}]{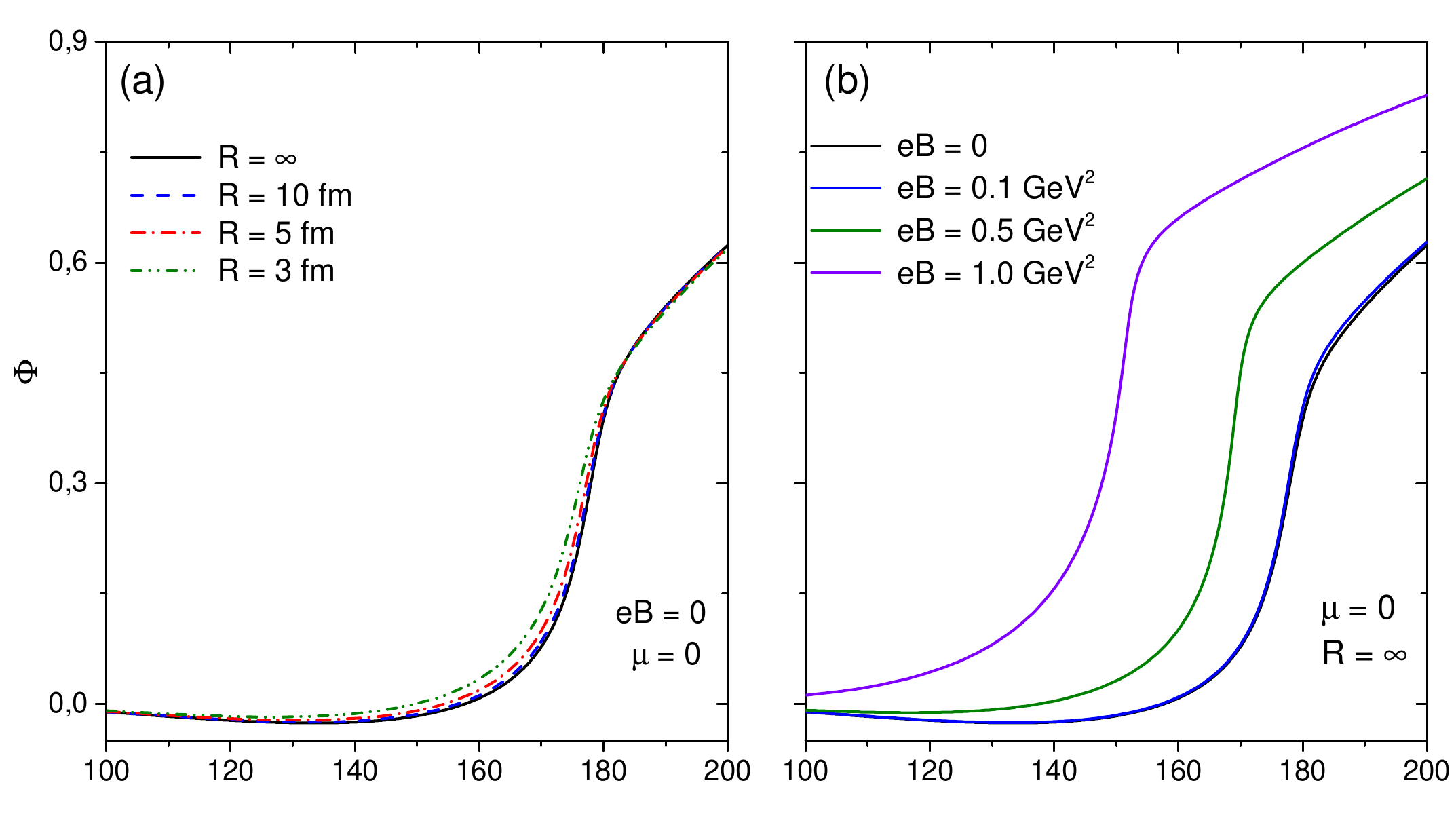} 
\includegraphics[width=0.80\textwidth,trim={0cm 1.20cm 0cm 0.0cm}]{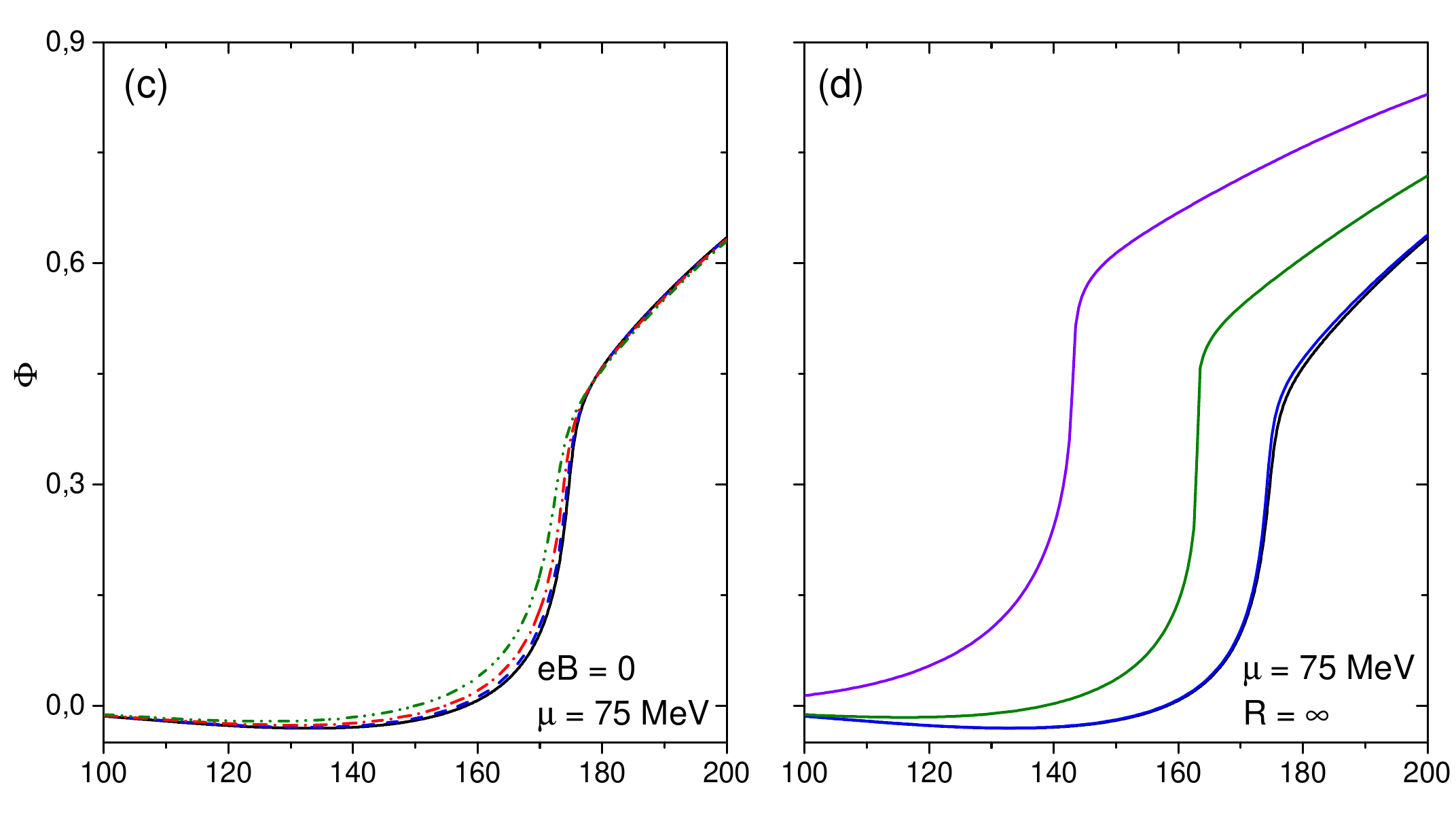} 
\includegraphics[width=0.80\textwidth,trim={0cm 0.00cm 0cm 0.0cm}]{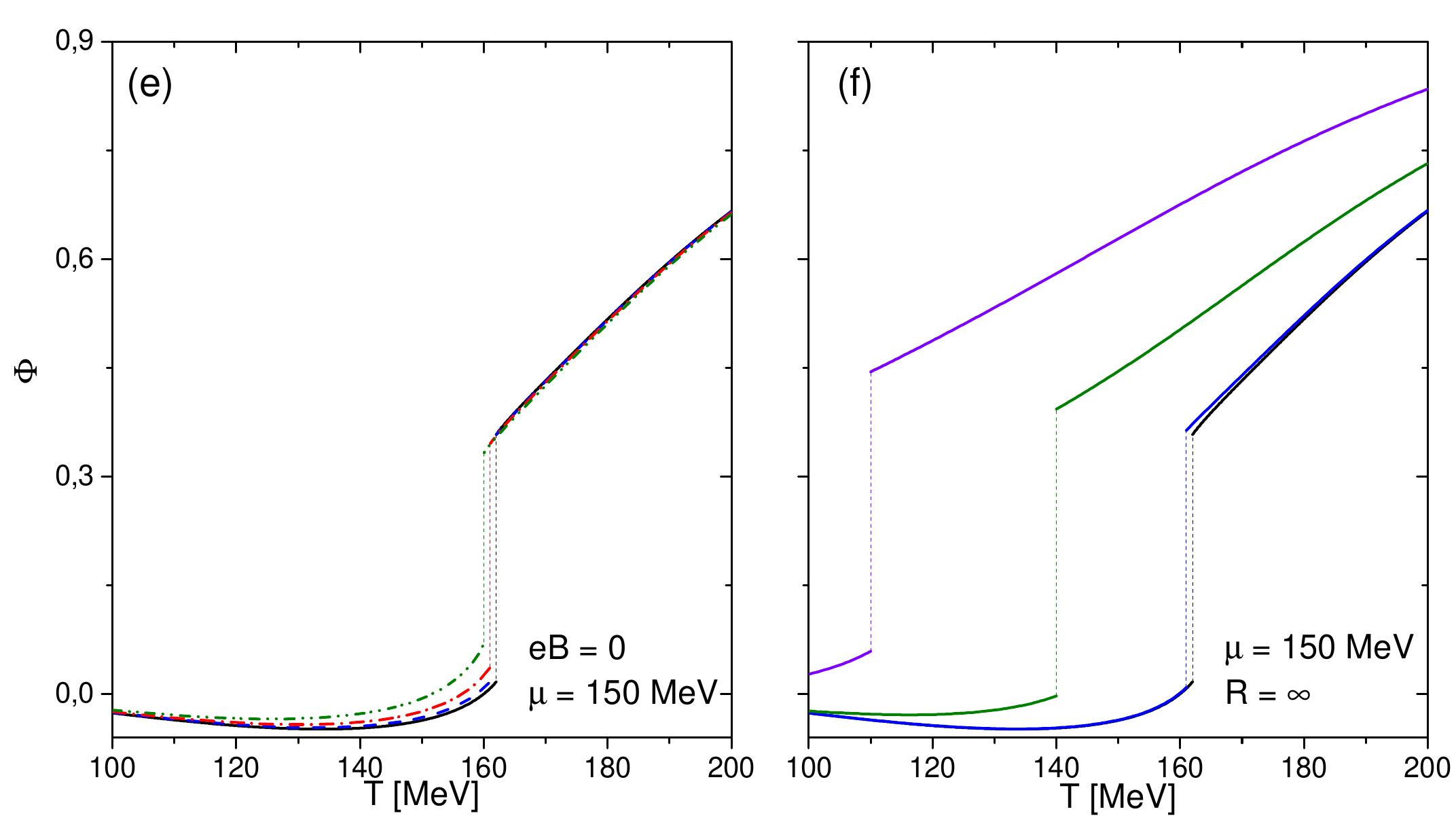} 
\caption{Traced Polyakov loop $\Phi$ as a function of the temperature $T$.
Left column: finite-size effects at $eB=0$ for several droplet radii.
Right column: bulk limit for different magnetic field strengths.
Rows correspond to different values of $\mu$.
The slightly negative values of $\Phi$ visible at low temperatures in some
panels are an artifact of the polynomial parametrization of the Polyakov-loop
potential (see discussion in the text).
\itemuno{The pseudocritical temperature extracted from $\chi_\Phi$ coincides with that
obtained from $\chi_{\rm ch}$ for all values of $R$ and $eB$ considered (see
Sec.~\ref{sec:formalism} for discussion).} }
\label{PHIvsT}
\end{figure}

\subsection{Deconfinement order parameter: the Polyakov loop}
\label{subsec:polyakov_results}

Figure~\ref{PHIvsT} displays the traced Polyakov loop $\Phi$ as a function
of temperature, following the same panel layout as
Fig.~\ref{fig:sigma_panels}.

The behavior of $\Phi$ mirrors closely that of $\bar\sigma$: the three
mechanisms identified in the preceding subsection---finite-size suppression,
the dual role of the magnetic field, and the steepening induced by
$\mu$---manifest in the same way in the deconfinement sector. Reducing the
droplet radius shifts the rise of $\Phi$ to lower temperatures
[panels~(a),~(c),~(e)], with the effect barely noticeable for $R=10$~fm
but clearly visible for $R=3$~fm. In the bulk limit, increasing $eB$
produces both an enhancement of the transition sharpness and a reduction
of the pseudocritical temperature [panels~(b),~(d),~(f)], consistently
with the IMC effect. At $\mu=150$~MeV and strong fields [panel~(f)],
$\Phi$ exhibits an abrupt jump, confirming the first-order character of
the transition already identified through $\bar\sigma$.

A technical remark is in order: in some panels $\Phi$ takes small negative
values at the lowest temperatures. This is a known artifact of the
polynomial parametrization of the Polyakov-loop potential, which does not
enforce $\Phi\geq 0$ strictly, and has no consequence for the transition
region. Likewise, the slight increase of $\Phi$ with $eB$ at low $T$
reflects the sensitivity of this quantity to the effective quark--gluon
coupling in the model, and we refrain from attributing a direct physical
meaning to this low-temperature behavior.


\begin{figure}[tb]
\centering
\includegraphics[width=0.80\textwidth,trim={0cm 1.20cm 0cm 1.0cm}]{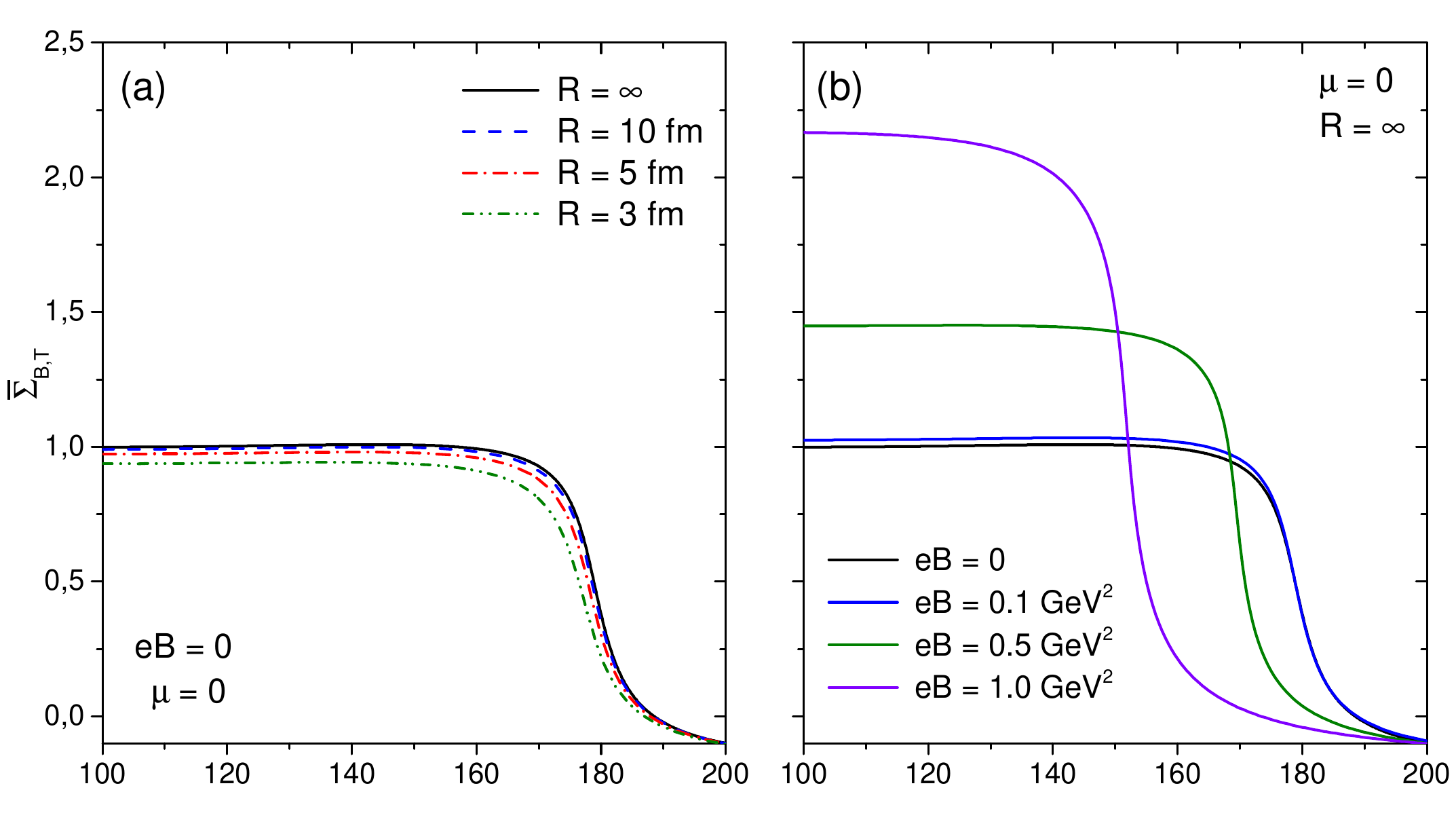} 
\includegraphics[width=0.80\textwidth,trim={0cm 1.20cm 0cm 0.0cm}]{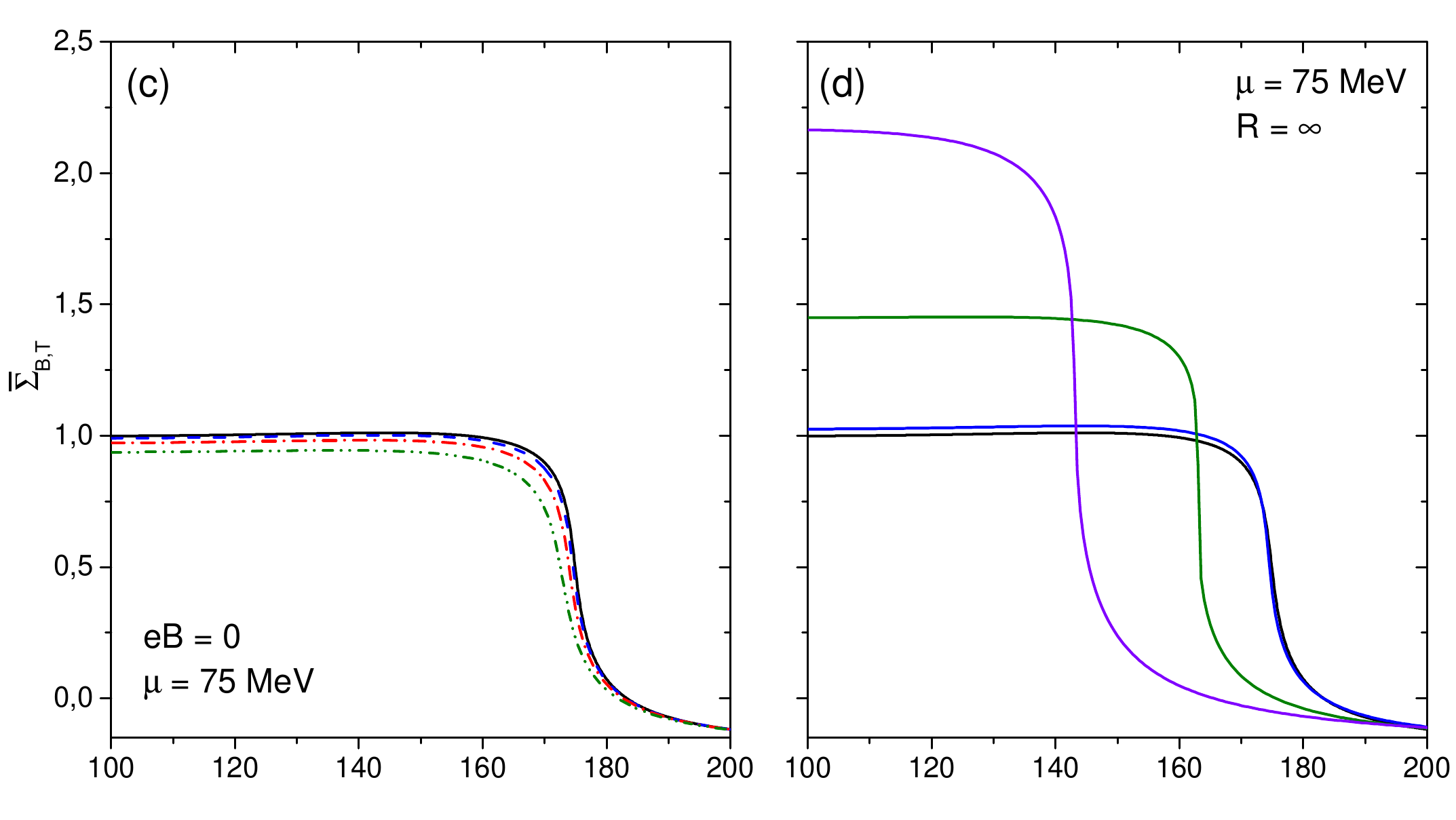} 
\includegraphics[width=0.80\textwidth,trim={0cm 0.00cm 0cm 0.0cm}]{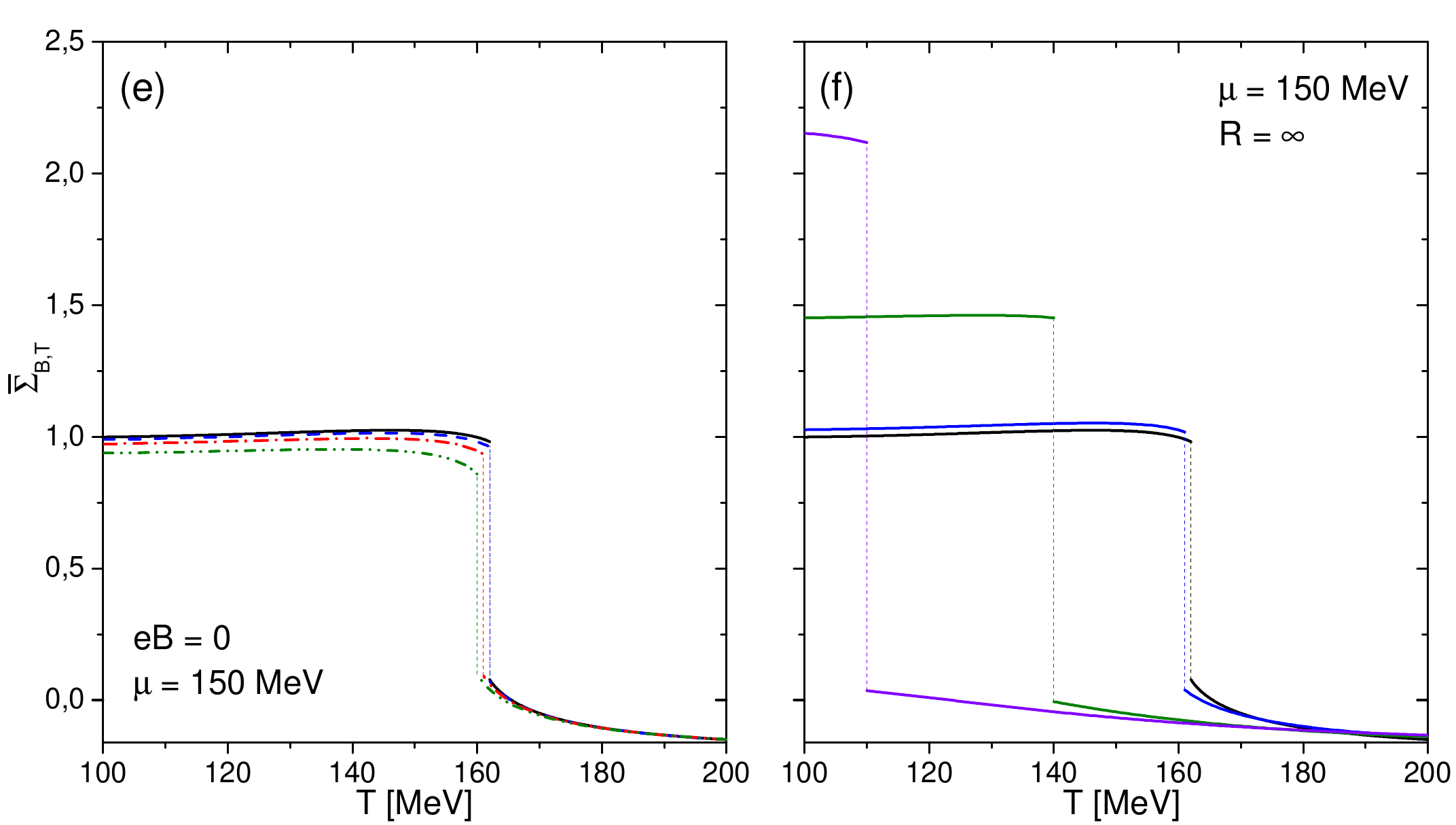} 
\caption{Normalized flavor-averaged condensate $\bar\Sigma_{B,T}$ as a
function of the temperature $T$.
Left column: finite-size effects at $eB=0$ for several droplet radii
($R=\infty$, $10$, $5$, and $3$~fm).
Right column: bulk limit ($R=\infty$) for different magnetic field strengths
($eB=0$, $0.1$, $0.5$, and $1.0$~GeV$^{2}$).
Rows correspond to $\mu=0$ [panels~(a) and~(b)], $\mu=75$~MeV
[panels~(c) and~(d)], and $\mu=150$~MeV [panels~(e) and~(f)].
The enhancement of $\bar\Sigma_{B,T}$ above unity at low $T$ in the right
column is a manifestation of magnetic catalysis.}
\label{condvsT}
\end{figure}

\subsection{Normalized flavour- averaged condensate $\bar\Sigma_{B,T}$}
\label{subsec:condensate_results}

We also analyze the normalized flavor-averaged condensate $\bar\Sigma_{B,T}$
defined in Sec.~\ref{sec:formalism}, which is normalized to unity at
$(T,eB)=(0,0)$ and decreases toward zero as chiral symmetry is restored.
Figure~\ref{condvsT} presents $\bar\Sigma_{B,T}(T)$ following the same
panel layout as in the preceding figures.

The behavior of $\bar\Sigma_{B,T}$ is fully consistent with the trends
already identified for $\bar\sigma$ and $\Phi$. Finite-size effects reduce
the low-temperature plateau and shift the transition to lower $T$, with the
suppression most visible for $R=3$~fm [left column]. The magnetic field
enhances $\bar\Sigma_{B,T}$ at low temperatures well above unity---reaching
$\sim 1.5$ at $eB=0.5$~GeV$^{2}$ and $\sim 2.0$ at
$eB=1.0$~GeV$^{2}$ for $\mu=0$ [panel~(b)]---providing a particularly
clear manifestation of magnetic catalysis in a quantity directly comparable
to lattice data. The pseudocritical temperature decreases with $eB$ and
the transition sharpens, becoming discontinuous at $\mu=150$~MeV for
$eB\geq 0.5$~GeV$^{2}$ [panel~(f)].

The concordance among Figs.~\ref{fig:sigma_panels},~\ref{PHIvsT},
and~\ref{condvsT} confirms the internal consistency of the calculation:
finite-size suppression, IMC-driven reduction of $T_{\rm pc}$, and magnetic
catalysis at low $T$ manifest uniformly across the mean field, the
Polyakov loop, and the normalized chiral condensate.


\begin{figure}[tb]
\centering
\includegraphics[width=0.80\textwidth,trim={0cm 1.20cm 0cm 1.0cm}]{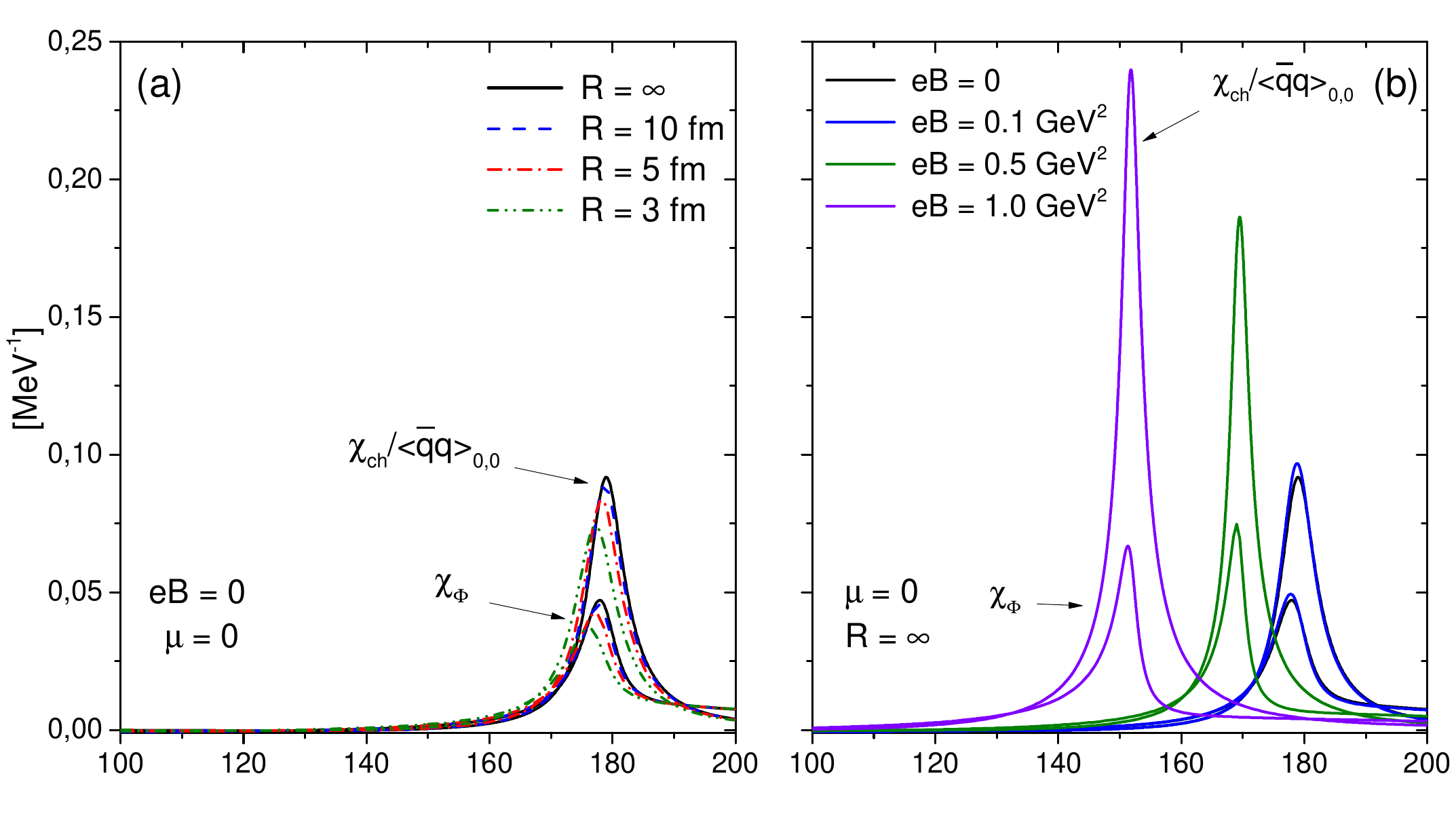} 
\includegraphics[width=0.80\textwidth,trim={0cm 0.00cm 0cm 0.0cm}]{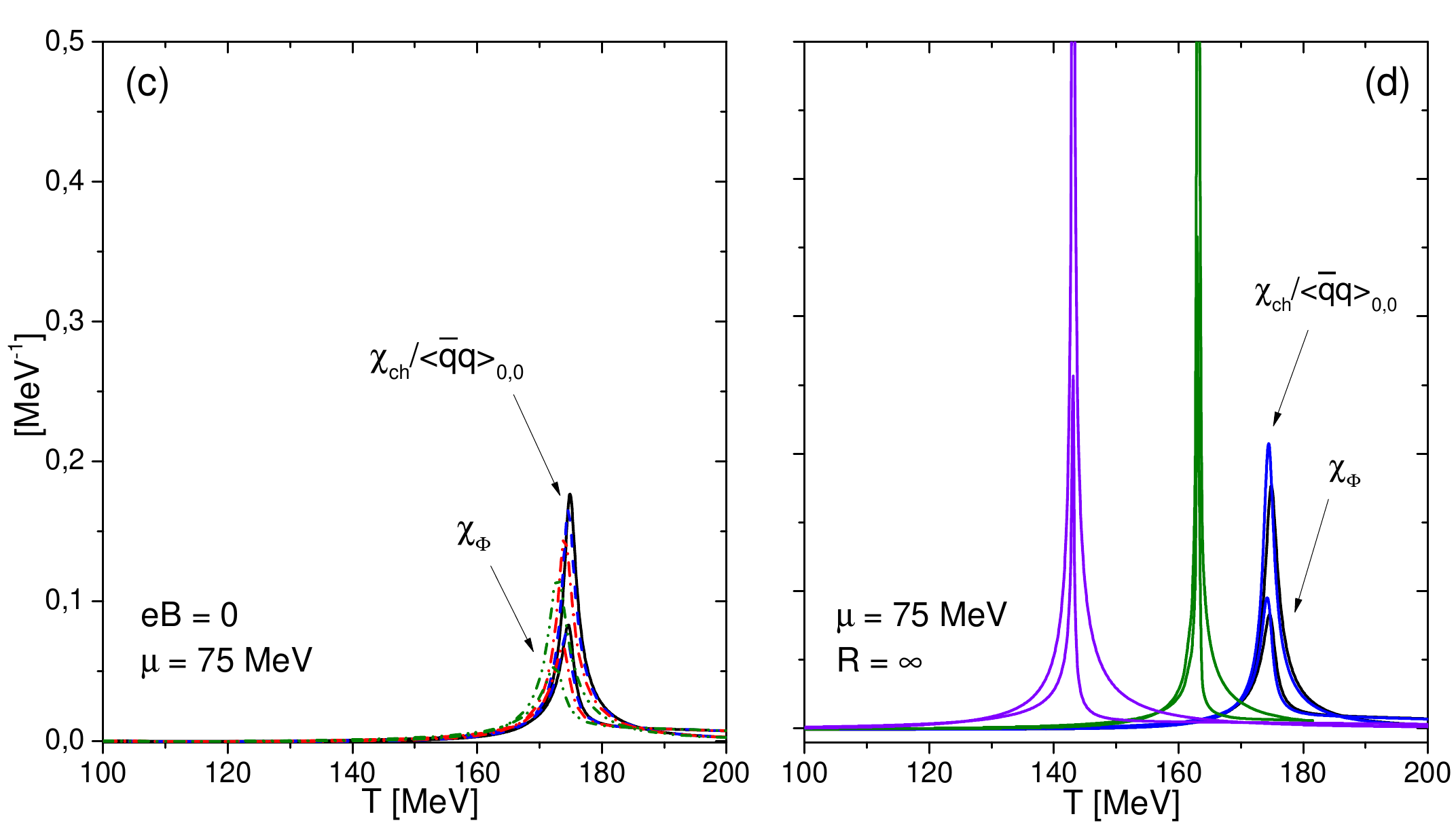} 
\caption{Chiral susceptibility $\chi_{\rm ch}/\langle\bar qq\rangle_{0,0}$
(solid curves) and Polyakov-loop susceptibility $\chi_\Phi$ (dashed curves)
as functions of the temperature $T$.
Left column: finite-size effects at $eB=0$ for several droplet radii
($R=\infty$, $10$, $5$, and $3$~fm).
Right column: bulk limit ($R=\infty$) for different magnetic field strengths
($eB=0$, $0.1$, $0.5$, and $1.0$~GeV$^{2}$).
The upper row corresponds to $\mu=0$ [panels~(a) and~(b)] and the lower row
to $\mu=75$~MeV [panels~(c) and~(d)].
\itemuno{The peaks of $\chi_{\rm ch}$ and $\chi_\Phi$ coincide within numerical
accuracy for all values of $R$ and $eB$ considered, with the caveats
discussed in Sec.~\ref{sec:formalism}.}}
\label{suscvsT}
\end{figure}

\subsection{Chiral and deconfinement susceptibilities}
\label{subsec:suscept_results}

The pseudocritical temperatures discussed above are determined
quantitatively from the peaks of $\chi_{\rm ch}$ and $\chi_\Phi$, defined
in Eq.~(\ref{susceptibilities}). Figure~\ref{suscvsT} shows both
susceptibilities as functions of $T$ for $\mu=0$ and $75$~MeV.

In all panels the peaks of $\chi_{\rm ch}$ and $\chi_\Phi$ coincide within
numerical accuracy, confirming the simultaneous character of the chiral and
deconfinement transitions established in the preceding subsections. \itemcuatro{This coincidence should be interpreted within the scope of the present implementation, where finite-size corrections act directly on the fermionic sector while $\mathcal U(\Phi,T)$ is kept in its bulk form. Consequently, the finite-$R$ behavior of $\Phi$ is induced indirectly through its coupling to the quark sector, and the persistence of the coincident peaks should not be viewed as an independent test of finite-volume gluonic dynamics.}

Reducing
the droplet radius shifts the peaks to lower temperatures and slightly
reduces their height [left column], reflecting the smoothing of the crossover
by finite-size effects. Increasing $eB$ in the bulk limit [right column]
also shifts the peaks to lower $T$---providing a direct quantitative measure
of IMC---while making them markedly taller and narrower, signaling the
approach to first-order behavior. At $\mu=75$~MeV the peaks are roughly
twice as tall as at $\mu=0$ ($\sim 0.4$--$0.5$ vs.\
$\sim 0.2$--$0.25$~MeV$^{-1}$), and the sharpening with $eB$ is even more
pronounced: at $eB=1.0$~GeV$^{2}$ [panel~(d)] the peak is very sharp,
indicating proximity to the critical end point.

These susceptibilities provide the pseudocritical temperatures used to
construct the crossover lines in the phase diagram presented in the next section.


\begin{figure}[tb]
\centering
\includegraphics[width=0.80\textwidth,trim={0cm 1.20cm 0cm 1.0cm}]{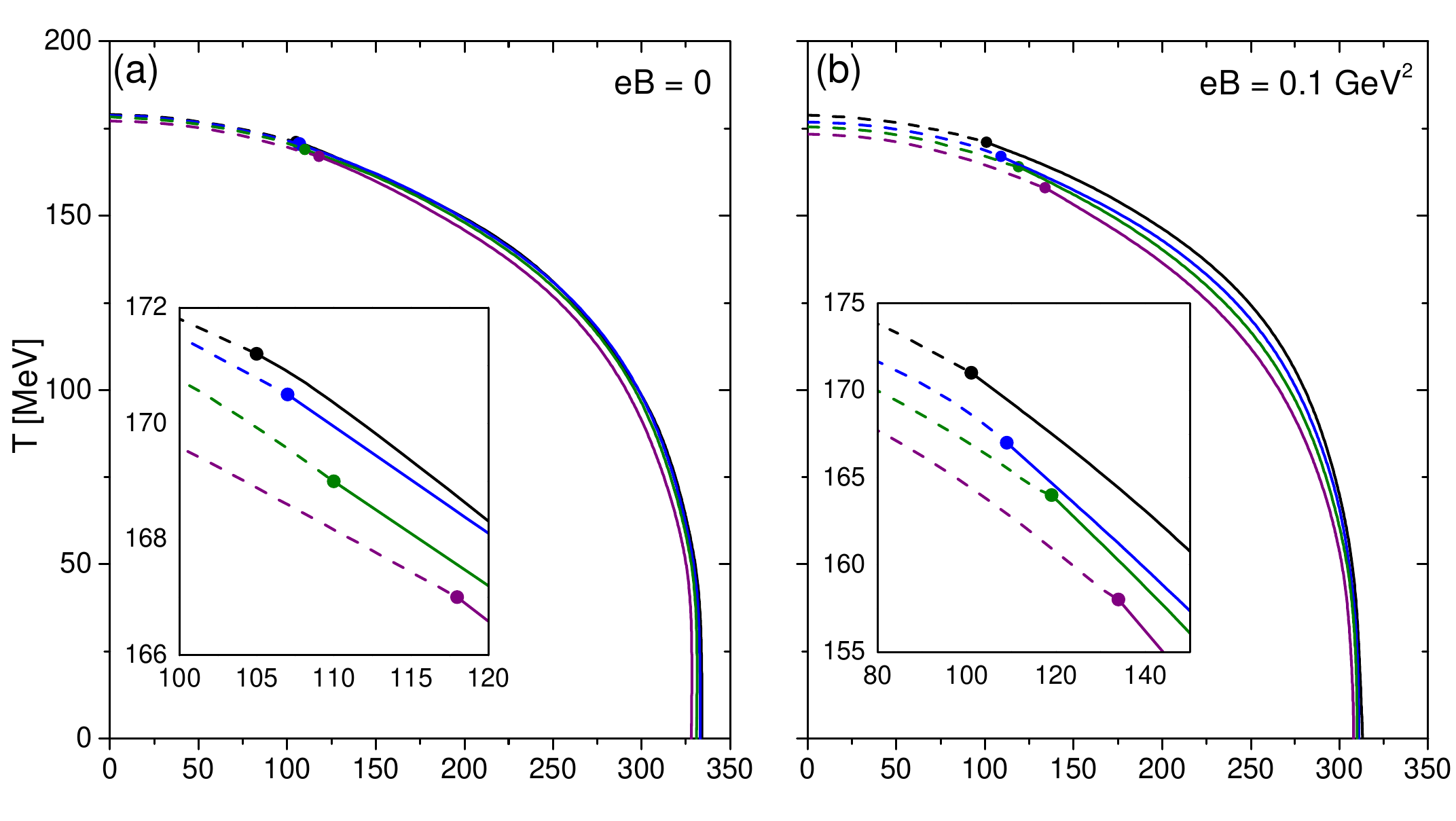} 
\includegraphics[width=0.80\textwidth,trim={0cm 0.00cm 0cm 0.0cm}]{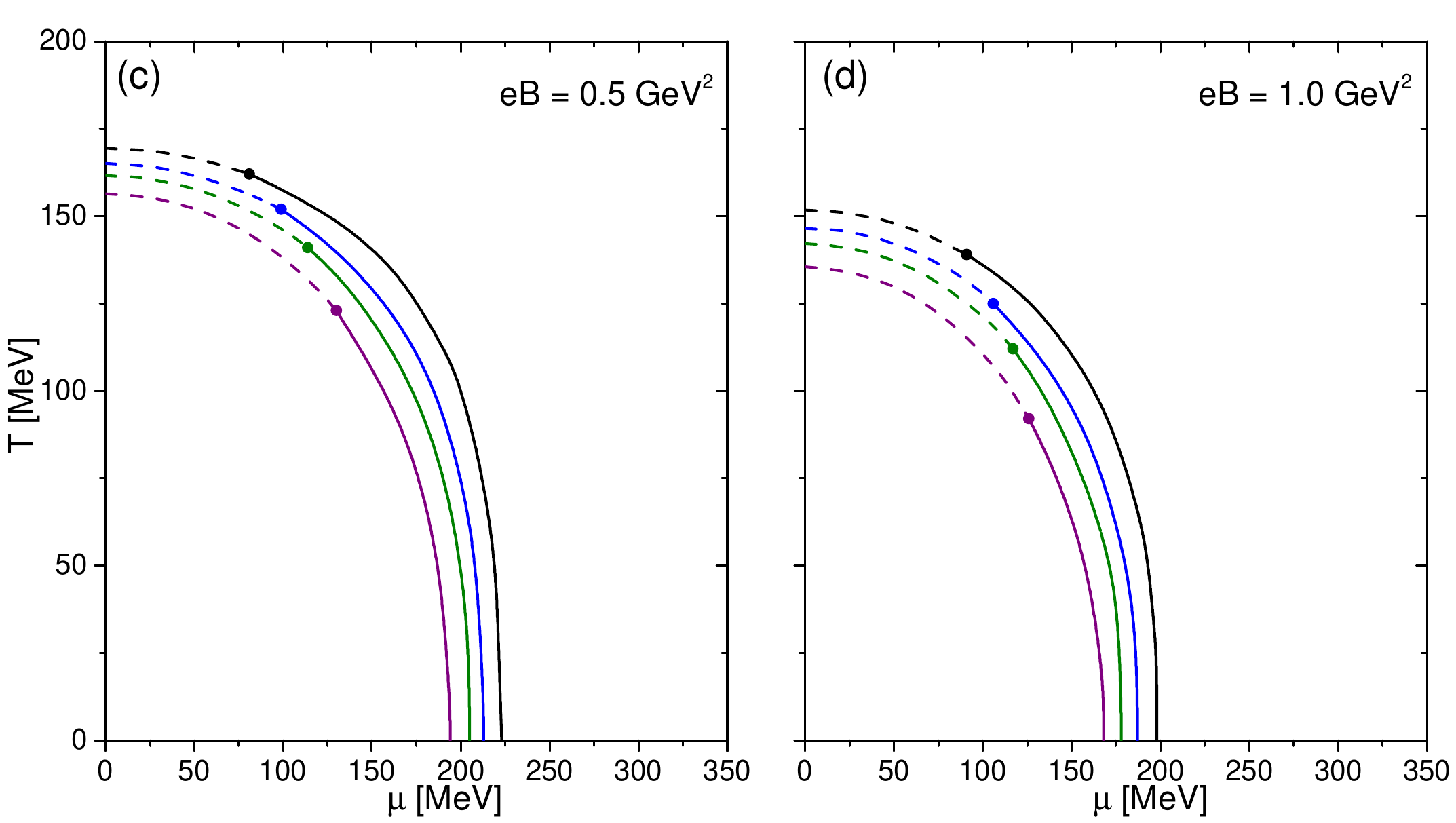}
\caption{Phase diagram in the $T$--$\mu$ plane for different magnetic field
strengths: $eB=0$ [panel~(a)], $0.1$~GeV$^{2}$ [panel~(b)],
$0.5$~GeV$^{2}$ [panel~(c)], and $1.0$~GeV$^{2}$ [panel~(d)]. In each panel,
solid lines denote the first-order transition and dashed lines denote the
crossover, for several droplet radii: $R=\infty$ (black), $10$~fm (blue),
$5$~fm (green), and $3$~fm (violet). Filled symbols mark the location of the
 CEP. At $eB=0$, the CEP coordinates are approximately
$(\mu_{\rm CEP},T_{\rm CEP})=(105,171)$~MeV for the bulk limit,
$(107,170)$~MeV for $R=10$~fm, $(110,169)$~MeV for $R=5$~fm, and
$(118,167)$~MeV for $R=3$~fm. Insets in panels~(a) and~(b) provide a
magnified view of the CEP region.}
\label{DDF}
\end{figure}

\section{Phase diagram in the $T$--$\mu$ plane}
\label{sec:phase_diagram}

We now assemble the information extracted from the order parameters and
susceptibilities into the phase diagram in the $T$--$\mu$ plane. Given that
the peaks of the chiral and Polyakov-loop susceptibilities occur at nearly
the same temperature for all values of $\mu$, $eB$, and $R$ explored, we
adopt the chiral susceptibility $\chi_{\rm ch}$ to define the pseudocritical
crossover temperature. In the first-order regime, the transition temperature
is identified by the Maxwell construction (equal depth of the two minima of
the thermodynamic potential), and the CEP is located as the boundary between
the crossover and first-order branches. The resulting pseudocritical temperatures 
at $\mu=0$ and CEP coordinates are collected in Tables~\ref{tab:T_C} 
and~\ref{tab:CEP_coordinates}, respectively.

Figure~\ref{DDF} presents the phase diagram for $eB=0$, $0.1$, $0.5$, and
$1.0$~GeV$^{2}$, each panel showing curves for several droplet radii. In
all cases the phase structure consists of a first-order transition line
(solid curves) meeting a crossover line (dashed curves) at a CEP (filled
symbol), with the crossover corresponding simultaneously to partial chiral
restoration and deconfinement.

The two main trends that emerge from the figure and the tables can be stated
concisely. First, at fixed system size, increasing $eB$ lowers the crossover
temperature at low $\mu$ (inverse magnetic catalysis) while strengthening the
first-order transition at higher densities. Table~\ref{tab:T_C} shows that in
the bulk limit $T_{\rm pc}(\mu=0)$ drops from $179$~MeV at $eB=0$ to
$152$~MeV at $eB=1.0$~GeV$^{2}$, a reduction of $\sim 27$~MeV. The CEP
temperature follows the same trend: $T_{\rm CEP}$ decreases from $171$~MeV at
$eB=0$ to $139$~MeV at $eB=1.0$~GeV$^{2}$ in the bulk
(Table~\ref{tab:CEP_coordinates}). This monotonic decrease of $T_{\rm CEP}$
with the magnetic field is opposite to the behavior reported in most studies
based on local NJL/PNJL
models~\citep{Avancini:2012ee,Ferrari:2012yw,Costa:2015bza}, and we attribute
it to the natural emergence of IMC in the nonlocal framework
as reported in Ref.~\citep{dumm2017strong} where it arises from the momentum 
dependence of the quark self-energy without ad hoc modifications.

Second, at fixed $eB$, decreasing the system size shifts the CEP toward
higher $\mu$ and lower $T$---and the magnitude of this shift grows
dramatically with the magnetic field. At $eB=0$ [panel~(a)] the effect is
subtle: between $R=\infty$ and $R=3$~fm the CEP moves by only
$\sim 12$~MeV in $\mu$ and $\sim 4$~MeV in $T$, and the crossover curves
for different radii are barely distinguishable except in the inset. At
$eB=0.1$~GeV$^{2}$ [panel~(b)] the separation begins to grow, with the CEP
for $R=3$~fm already shifted to $(\mu,T)\approx(134,158)$~MeV compared with
$(101,171)$~MeV in the bulk---a displacement of $33$~MeV in $\mu$ and
$13$~MeV in $T$. At $eB=0.5$~GeV$^{2}$ [panel~(c)] the CEP positions for
different radii are clearly resolved without magnification, and the spread
reaches $\sim 49$~MeV in $\mu$ and $\sim 39$~MeV in $T$ between the bulk
and $R=3$~fm. At $eB=1.0$~GeV$^{2}$ [panel~(d)] the finite-size effects
are most dramatic: the $R=3$~fm CEP lies at $(126,92)$~MeV, some $35$~MeV
higher in $\mu$ and $47$~MeV lower in $T$ than the bulk value of
$(91,139)$~MeV. In this regime the phase diagrams for different radii differ
qualitatively across the entire $T$--$\mu$ plane, not merely near the CEP.

This amplification of finite-size effects by the magnetic field can be
understood from the interplay between Landau quantization and geometric
confinement. The dimensional reduction associated with the lowest Landau
level concentrates the low-energy dynamics along the field direction,
enhancing the sensitivity to infrared physics. The MRE infrared cutoff
$\Lambda_{\rm IR}$, which precisely suppresses long-wavelength modes in the
LLL sector, therefore has a progressively larger impact on the phase
structure as $eB$ increases. The pseudocritical temperatures at $\mu=0$
confirm this picture quantitatively: Table~\ref{tab:T_C} shows that the
spread in $T_{\rm pc}$ between $R=\infty$ and $R=3$~fm increases from less than
$2$~MeV at $eB=0$ to more than $16$~MeV at $eB=1.0$~GeV$^{2}$.

\itemdos{
For \(R=3\) fm one has \(1/R\simeq 65\) MeV, while the corresponding MRE cutoff
for light quarks is \(\Lambda_{\rm IR}\simeq 52.5\) MeV. Thus, the cutoff acts
precisely in the momentum region where the truncated expansion is least
controlled. This does not invalidate the qualitative finite-size mechanism, but
it implies that the corresponding numerical shift of the CEP should be regarded
as prescription dependent. 
At \(B=0\), the infrared region affected by the cutoff is weighted by the three-dimensional phase-space measure \(p^2dp\), which suppresses the relative contribution of the lowest momenta to the thermodynamic integrals. Therefore, the cutoff modifies the thermodynamics but does not dominate the full momentum integration. In a strong magnetic field, however, Landau-level quantization reduces the effective dimensionality of the low-energy phase space. In particular, when the lowest Landau level becomes dominant, the infrared modes removed by the cutoff carry comparatively more weight. This explains why the finite-size shift is amplified at large \(eB\).

Therefore, the trends found in this work---the displacement of the CEP toward
larger chemical potentials and lower temperatures with decreasing \(R\), and
the enhancement of this displacement by the magnetic field---are interpreted as
robust consequences of geometric infrared suppression and Landau-level
dimensional reduction. The precise numerical CEP coordinates, especially for
the limiting case \(R=3\) fm at large \(eB\), should instead be understood as
model- and prescription-dependent.
}

These results indicate that finite-size corrections, which are often
neglected in effective-model studies, can have a significant quantitative
impact on the predicted location of the CEP---particularly in the presence
of strong magnetic fields. This observation is relevant both for the
interpretation of heavy-ion collision experiments, where the QGP has a
spatial extent of a few femtometers, and for the microphysics of phase
conversion in the interior of compact stars, where quark matter droplets 
could nucleate in the presence of intense magnetic fields.

\begin{table}[tb]
\centering
\caption{Pseudocritical temperature $T_{\rm pc}$ in MeV at zero chemical
potential, for each combination of magnetic field strength $eB$ and droplet
radius $R$ explored in this work.}
\label{tab:T_C}
\begin{tabular}{l c c c c}
\hline\hline
& $R=\infty$ & $R=10$~fm & $R=5$~fm & $R=3$~fm \\
\hline
$eB=0$  & $179.1$ & $178.7$ & $178.3$ & $177.2$ \\
$eB=0.1$~GeV$^{2}$  & $178.8$ & $176.9$ & $175.5$ & $173.4$ \\
$eB=0.5$~GeV$^{2}$  & $169.5$ & $165.1$ & $161.6$ & $156.4$ \\
$eB=1.0$~GeV$^{2}$  & $151.8$ & $146.5$ & $142.2$ & $135.5$ \\
\hline\hline
\end{tabular}
\label{Tcmu0}
\end{table}

\begin{table}[tb]
\centering
\caption{Coordinates of the critical end point
$(\mu_{\rm CEP},T_{\rm CEP})$ in MeV, for all values of $eB$ and $R$ considered.}
\label{tab:CEP_coordinates}
\begin{tabular}{l c c c c}
\hline\hline
& $R=\infty$ & $R=10$~fm & $R=5$~fm & $R=3$~fm \\
\hline
$eB=0$  & $(106,\,171)$ & $(107,\,170)$ & $(110,\,169)$ & $(118,\,167)$ \\
$eB=0.1$~GeV$^{2}$  & $(101,\,171)$ & $(109,\,167)$ & $(119,\,164)$ & $(134,\,158)$ \\
$eB=0.5$~GeV$^{2}$  & $(\phantom{0}81,\,162)$ & $(\phantom{0}99,\,152)$ & $(114,\,141)$ & $(130,\,123)$ \\
$eB=1.0$~GeV$^{2}$  & $(\phantom{0}91,\,139)$ & $(106,\,125)$ & $(117,\,112)$ & $(126,\,\phantom{0}92)$ \\
\hline\hline
\end{tabular}
\end{table}

\section{Summary and conclusions}
\label{sec:conclusions}

We have studied the QCD phase diagram at finite temperature and chemical
potential, in the presence of an external static and uniform magnetic field,
incorporating finite-size effects through the  MRE formalism. Our analysis 
has been carried out in the framework of a two-flavor nlPNJL model, which has
the distinctive feature of reproducing the IMC effect observed in lattice-QCD 
calculations, without the need for ad hoc modifications to the model parameters.

The main results of this work can be summarized as follows:

(i) Previous studies in the bulk limit have shown that the nonlocal PNJL
model naturally exhibits magnetic catalysis at low temperatures and inverse
magnetic catalysis near the pseudocritical
temperature~\citep{Pagura:2016pwr,dumm2017strong,Carlomagno:2023clk}. In the
present work we demonstrate that these features remain robust in the
presence of finite-size effects: for all system sizes considered and at
finite chemical potential, the chiral condensate increases with the magnetic
field at low temperatures, while the pseudocritical temperature $T_{\rm pc}$
decreases monotonically with $eB$.

\itemuno{
(ii) The crossover transitions, characterized respectively by the chiral susceptibility $\chi_{\rm ch}$ and
the Polyakov-loop susceptibility $\chi_\Phi$, occur simultaneously
throughout the explored parameter space. This coincidence, already known in the bulk limit, is shown here to persist in
magnetized quark matter after the inclusion of finite-size effects, down to
$R=3$~fm. Since the MRE corrections are applied only to the fermionic
sector while $\mathcal U(\Phi,T)$ is kept in its bulk form---a standard
choice in PNJL studies of finite-volume effects, supported by the
hierarchy between the gluonic correlation length and the droplet
radii considered here---the $R$-dependence of $\Phi$ is inherited from
its coupling to the quarks, and a fully independent test of this
coincidence at finite volume would require a finite-size formulation of
the pure-gauge sector.
}

(iii) The qualitative structure of the phase diagram---consisting of a
crossover region at low $\mu$ and a first-order transition line at higher
$\mu$, separated by a CEP---is maintained for all values
of $eB$ and $R$ explored (see Table~\ref{tab:CEP_coordinates} for a complete
listing of the CEP coordinates). In the bulk limit, the CEP temperature
decreases monotonically with the magnetic field, from
$T_{\rm CEP}\simeq 171$~MeV at $eB=0$ to $T_{\rm CEP}\simeq 140$~MeV at
$eB=1$~GeV$^{2}$, while the chemical potential $\mu_{\rm CEP}$ remains in a
relatively narrow range ($\sim 80$--$105$~MeV). This behavior is opposite to
that found in most studies based on local NJL/PNJL
models~\citep{Avancini:2012ee,Ferrari:2012yw,Costa:2013zca,Costa:2015bza,Ferreira:2017wtx},
and we attribute this qualitative difference to the natural emergence of IMC
in the nonlocal framework. Indeed, in local models the behavior of
$T_{\rm CEP}$ with the magnetic field is significantly modified when a
$B$-dependent effective coupling is introduced to mimic
IMC~\citep{Costa:2015bza}.

(iv) Finite-size effects shift the CEP toward higher chemical potentials and
lower temperatures. At $eB=0$, this shift is modest: from
$(\mu_{\rm CEP},T_{\rm CEP})\approx(105,171)$~MeV in the bulk to
$(118,167)$~MeV for $R=3$~fm. However, the sensitivity of the phase
structure to the system size is significantly amplified by the magnetic field.
At $eB=1$~GeV$^{2}$, the CEP positions for different radii are well separated
across the entire $T$--$\mu$ plane, reflecting the interplay between
the dimensional reduction induced by Landau quantization and the infrared
mode suppression imposed by the MRE formalism.

The physical origin of this amplification can be understood as follows.
In a strong magnetic field, the transverse quark motion is quantized into
discrete Landau levels separated by gaps of order $|q_f B|$. As the field
strength increases, the higher levels become energetically suppressed and
the low-energy dynamics is increasingly dominated by the LLL, 
in which only the longitudinal momentum component $p_3$
remains as a continuous degree of freedom. The system thus undergoes an
effective dimensional reduction from three to one momentum-space dimension.
In a bulk system ($R\to\infty$), this concentration of spectral weight in
the LLL modifies the phase structure but preserves the full infrared
spectrum along $p_3$. When the system is confined to a finite volume, the
MRE formalism introduces an effective infrared cutoff $\Lambda_{\rm IR}$
that suppresses long-wavelength modes. In the absence of a magnetic field,
this cutoff removes modes from a three-dimensional momentum space, where
the affected states represent a small fraction of the total density of
states, and the thermodynamic impact is correspondingly modest. In
contrast, when the magnetic field is strong enough to enforce
LLL dominance, the infrared cutoff acts on the single remaining continuous
momentum direction, removing a proportionally much larger fraction of the
states that govern the low-energy thermodynamics. It is this combination of
dimensional reduction and infrared suppression that makes the phase
structure---and in particular the CEP location---increasingly sensitive to
the system size as $eB$ grows.

It is important to distinguish which aspects of these results are expected
to be robust and which may depend on the specific implementation of
finite-size effects. The qualitative trends---namely, that confinement to a
finite volume shifts the CEP toward higher $\mu$ and lower $T$, and that
this shift is amplified by the magnetic field---rest on general physical
grounds: the suppression of infrared modes by geometric confinement and the
dimensional reduction induced by Landau quantization are well-established
features that do not depend on the MRE formalism. The quantitative magnitude 
of the effect, however, is more sensitive to the details of the finite-size
treatment.  The MRE models the boundary as a sharp spherical surface with 
specific boundary conditions---an idealization of the actual diffuse 
interface---and, being a semiclassical expansion in powers of $1/R$, 
it may become less accurate at the smallest radii.

(v) Finite-size effects also modify the mean field and the chiral
condensate: reducing the system radius suppresses the chiral condensate at
temperatures below the transition and shifts the crossover to lower
temperatures, smoothing the transition. At a fixed point in the $T$--$\mu$
plane, these effects act cumulatively with those of the magnetic field and
the chemical potential: reducing $R$, increasing $eB$, or raising $\mu$ can
each push the system from the crossover regime into the first-order region,
as the CEP shifts accordingly. However, as noted in point~(iii), the overall
topology of the phase diagram---crossover at low $\mu$, first order at
higher $\mu$, separated by a CEP---is preserved for all parameter
combinations explored.

Our analysis has been restricted to two flavors; however, the values of
$\mu_{\rm CEP}$ obtained for all combinations of $eB$ and $R$ lie well
below the strange quark threshold. In addition,
Ref.~\citep{Carlomagno:2023clk} showed that IMC at finite $\mu$ is robust
under moderate variations of the model parameters in the bulk limit.  
We therefore expect the qualitative trends reported here to survive 
the inclusion of strangeness, although a richer phase structure could 
appear at higher densities~\citep{Ferreira:2017wtx}.

\itemtres{
The findings of this work may have implications for relativistic heavy-ion
collisions, where the quark-gluon plasma formed at the earliest stages is finite
in size and exposed to intense, rapidly evolving magnetic fields. Since the
magnitude and lifetime of these fields depend on the collision energy,
centrality, and electromagnetic response of the medium, the lower and
intermediate values of \(eB\) considered here provide the more conservative
phenomenological range. By contrast, \(eB=1~\mathrm{GeV}^2\) should be regarded
as an extreme theoretical case, useful to probe the model behavior rather than
as a quantitatively realistic heavy-ion scenario.

Within this interpretation, the large-field results show how Landau-level
quantization, especially the increasing dominance of the lowest Landau level,
amplifies the sensitivity of the CEP to finite-size infrared suppression.
Therefore, the robust conclusion is not the precise numerical displacement of
the CEP at \(eB=1~\mathrm{GeV}^2\), but the qualitative trend that magnetic
fields enhance finite-size effects in the phase diagram. This interplay between
finite geometry and magnetic quantization provides a useful perspective for
assessing how nonbulk effects may shape the QCD transition region in small,
magnetized systems.
}

\vspace{6pt}

\authorcontributions{Conceptualization, methodology, analysis and writing, AGG and GL; software and visualization, AGG and SAF. All authors have read and agreed to the published version of the manuscript.}

\funding{GL acknowledges the partial financial support from the Brazilian agency CNPq (grant 316844/2021-7). AGG acknowledges the financial support from CONICET under Grant No. PIP 22-24 11220210100150CO, and the National University of La Plata (Argentina), Project No. X960.}

\dataavailability{The raw data supporting the conclusions of this article will be made available by the authors on request.}


\conflictsofinterest{The authors declare no conflict of interest.}

\abbreviations{Abbreviations}{
The following abbreviations are used in this manuscript:\\

\noindent 
\begin{tabular}{@{}ll}

PNJL & Polyakov Nambu Jona-Lasinio \\
MRE & multiple reflection expansion  \\
nlPNJL & non-local Polyakov Nambu Jona-Lasinio \\
CEP & critical endpoint\\
QCD & Quantum Chromodynamics\\
LQCD & Lattice Quantum Chromodynamics\\
NJL & Nambu Jona-Lasinio \\
MC & magnetic catalysis \\
IMC & inverse magnetic catalysis \\
MFA & mean field approximation \\
LLL & lowest Landau level 
\end{tabular}
}

\begin{adjustwidth}{-\extralength}{0cm}
\reftitle{References}
\bibliography{TMUB}

@article{Madsen:1994vp,
    author = "Madsen, Jes",
    title = "{Shell model versus liquid drop model for strangelets}",
    eprint = "hep-ph/9407314",
    archivePrefix = "arXiv",
    doi = "10.1103/PhysRevD.50.3328",
    journal = "Phys. Rev. D",
    volume = "50",
    pages = "3328--3331",
    year = "1994"
}

@article{Balian:1970fw,
    author = "Balian, R. and Bloch, C.",
    title = "{Distribution of eigenfrequencies for the wave equation in a finite domain. 1. Three-dimensional problem with smooth boundary surface}",
    doi = "10.1016/0003-4916(70)90497-5",
    journal = "Annals Phys.",
    volume = "60",
    pages = "401--447",
    year = "1970"
}

@article{Abreu2006,
  author        = {Abreu, L. M. and Gomes, M. and da Silva, A. J.},
  title         = {{Chiral symmetry breaking in a finite volume}},
  journal       = {Phys. Lett. B},
  volume        = {642},
  pages         = {551--562},
  year          = {2006},
  doi           = {10.1016/j.physletb.2006.10.015},
  eprint        = {hep-th/0610111},
  archivePrefix = {arXiv}
}

@article{Yasui2006,
  author  = {Yasui, Shigehiro and Hosaka, Atsushi},
  title   = {{Finite size effect on chiral symmetry breaking in the Nambu--Jona-Lasinio model}},
  journal = {Phys. Rev. D},
  volume  = {74},
  pages   = {054036},
  year    = {2006},
  doi     = {10.1103/PhysRevD.74.054036}
}

@article{Abreu2011,
  author        = {Abreu, L. M. and Malbouisson, A. P. C. and Malbouisson, J. M. C.},
  title         = {{Finite-size effects on the phase diagram of difermion condensates in two-dimensional four-fermion interaction models}},
  journal       = {Phys. Rev. D},
  volume        = {83},
  pages         = {025001},
  year          = {2011},
  doi           = {10.1103/PhysRevD.83.025001}
}

@article{Cristoforetti2010,
  author        = {Cristoforetti, Marco and Hell, Thomas and Klein, Bertram and Weise, Wolfram},
  title         = {{Finite-volume effects on phase transition in the Polyakov-loop extended Nambu--Jona-Lasinio model}},
  journal       = {Phys. Rev. D},
  volume        = {81},
  pages         = {114017},
  year          = {2010},
  doi           = {10.1103/PhysRevD.81.114017},
  eprint        = {1002.2336},
  archivePrefix = {arXiv},
  primaryClass  = {hep-ph}
}

@article{Bhattacharyya:2012rp,
    author = "Bhattacharyya, Abhijit and Deb, Paramita and Ghosh, Sanjay K. and Ray, Rajarshi and Sur, Subrata",
    title = "{Thermodynamic Properties of Strongly Interacting Matter in Finite Volume using Polyakov-Nambu-Jona-Lasinio Model}",
    eprint = "1212.5893",
    archivePrefix = "arXiv",
    primaryClass = "hep-ph",
    doi = "10.1103/PhysRevD.87.054009",
    journal = "Phys. Rev. D",
    volume = "87",
    number = "5",
    pages = "054009",
    year = "2013"
}

@article{Bhattacharyya:2014uxa,
    author = "Bhattacharyya, Abhijit and Ray, Rajarshi and Sur, Subrata",
    title = "{Fluctuation of strongly interacting matter in the Polyakov{\textendash}Nambu{\textendash}Jona-Lasinio model in a finite volume}",
    eprint = "1412.8316",
    archivePrefix = "arXiv",
    primaryClass = "hep-ph",
    doi = "10.1103/PhysRevD.91.051501",
    journal = "Phys. Rev. D",
    volume = "91",
    number = "5",
    pages = "051501",
    year = "2015"
}

@article{Tripolt2014,
    author = "Tripolt, Ralf-Arno and Braun, Jens and Klein, Bertram and Schaefer, Bernd-Jochen",
    title = "{Effect of fluctuations on the QCD critical point in a finite volume}",
    eprint = "1308.0164",
    archivePrefix = "arXiv",
    primaryClass = "hep-ph",
    doi = "10.1103/PhysRevD.90.054012",
    journal = "Phys. Rev. D",
    volume = "90",
    number = "5",
    pages = "054012",
    year = "2014"
}

@article{Pan2017,
    author = "Pan, Zan and Cui, Zhu-Fang and Chang, Chao-Hsi and Zong, Hong-Shi",
    title = "{Finite-volume effects on phase transition in the Polyakov-loop extended Nambu{\textendash}Jona-Lasinio model with a chiral chemical potential}",
    eprint = "1611.07370",
    archivePrefix = "arXiv",
    primaryClass = "hep-ph",
    doi = "10.1142/S0217751X17500671",
    journal = "Int. J. Mod. Phys. A",
    volume = "32",
    number = "13",
    pages = "1750067",
    year = "2017"
}

@article{Braun2011,
    author = "Braun, Jens and Klein, Bertram and Schaefer, Bernd-Jochen",
    title = "{On the Phase Structure of QCD in a Finite Volume}",
    eprint = "1110.0849",
    archivePrefix = "arXiv",
    primaryClass = "hep-ph",
    doi = "10.1016/j.physletb.2012.05.053",
    journal = "Phys. Lett. B",
    volume = "713",
    pages = "216--223",
    year = "2012"
}

@article{Abreu2019,
    author = "Abreu, L. M. and Corr{\^e}a, Emerson B. S. and Linhares, Cesar A. and Malbouisson, Adolfo P. C.",
    title = "{Finite-volume and magnetic effects on the phase structure of the three-flavor Nambu{\textendash}Jona-Lasinio model}",
    eprint = "1903.09249",
    archivePrefix = "arXiv",
    primaryClass = "hep-ph",
    doi = "10.1103/PhysRevD.99.076001",
    journal = "Phys. Rev. D",
    volume = "99",
    number = "7",
    pages = "076001",
    year = "2019"
}

@article{MataCarrizal2022,
    author = "Mata Carrizal, Nallaly Berenice and Valbuena Ord{\'o}{\~n}ez, Enrique and Garza Aguirre, Adri{\'a}n Jacob and Betancourt Sotomayor, Francisco Javier and Morones Ibarra, Jos{\'e} Rub{\'e}n",
    title = "{Effects of a Finite Volume in the Phase Structure of QCD}",
    doi = "10.3390/universe8050264",
    journal = "Universe",
    volume = "8",
    number = "5",
    pages = "264",
    year = "2022"
}

@article{CastanoYepes2022,
    author = "Casta{\~n}o-Yepes, Jorge David and Paniagua, Fernando Mart{\'\i}nez and Mu{\~n}oz-Vitelly, Victor and Ramirez-Gutierrez, Cristian Felipe",
    title = "{Volume effects on the QCD critical end point from thermal fluctuations within the super statistics framework}",
    eprint = "2208.06747",
    archivePrefix = "arXiv",
    primaryClass = "hep-ph",
    doi = "10.1103/PhysRevD.106.116019",
    journal = "Phys. Rev. D",
    volume = "106",
    number = "11",
    pages = "116019",
    year = "2022"
}

@article{Bernhardt2021,
    author = "Bernhardt, Julian and Fischer, Christian S. and Isserstedt, Philipp and Schaefer, Bernd-Jochen",
    title = "{Critical endpoint of QCD in a finite volume}",
    eprint = "2107.05504",
    archivePrefix = "arXiv",
    primaryClass = "hep-ph",
    doi = "10.1103/PhysRevD.104.074035",
    journal = "Phys. Rev. D",
    volume = "104",
    number = "7",
    pages = "074035",
    year = "2021"
}

@article{Kharzeev:2012ph,
    author = "Kharzeev, Dmitri E. and Landsteiner, Karl and Schmitt, Andreas and Yee, Ho-Ung",
    title = "{'Strongly interacting matter in magnetic fields': an overview}",
    eprint = "1211.6245",
    archivePrefix = "arXiv",
    primaryClass = "hep-ph",
    doi = "10.1007/978-3-642-37305-3_1",
    journal = "Lect. Notes Phys.",
    volume = "871",
    pages = "1--11",
    year = "2013"
}

@article{Andersen:2014xxa,
    author = "Andersen, Jens O. and Naylor, William R. and Tranberg, Anders",
    title = "{Phase diagram of QCD in a magnetic field: A review}",
    eprint = "1411.7176",
    archivePrefix = "arXiv",
    primaryClass = "hep-ph",
    doi = "10.1103/RevModPhys.88.025001",
    journal = "Rev. Mod. Phys.",
    volume = "88",
    pages = "025001",
    year = "2016"
}

@article{Miransky:2015ava,
    author = "Miransky, Vladimir A. and Shovkovy, Igor A.",
    title = "{Quantum field theory in a magnetic field: From quantum chromodynamics to graphene and Dirac semimetals}",
    eprint = "1503.00732",
    archivePrefix = "arXiv",
    primaryClass = "hep-ph",
    doi = "10.1016/j.physrep.2015.02.003",
    journal = "Phys. Rept.",
    volume = "576",
    pages = "1--209",
    year = "2015"
}

@article{Duncan:1992hi,
    author = "Duncan, Robert C. and Thompson, Christopher",
    title = "{Formation of very strongly magnetized neutron stars - implications for gamma-ray bursts}",
    doi = "10.1086/186413",
    journal = "Astrophys. J. Lett.",
    volume = "392",
    pages = "L9",
    year = "1992"
}

@article{Chatterjee:2014qsa,
    author = "Chatterjee, Debarati and Elghozi, Thomas and Novak, Jerome and Oertel, Micaela",
    title = "{Consistent neutron star models with magnetic field dependent equations of state}",
    eprint = "1410.6332",
    archivePrefix = "arXiv",
    primaryClass = "astro-ph.HE",
    doi = "10.1093/mnras/stu2706",
    journal = "Mon. Not. Roy. Astron. Soc.",
    volume = "447",
    pages = "3785",
    year = "2015"
}

@article{Skokov:2009qp,
    author = "Skokov, V. and Illarionov, A. Yu. and Toneev, V.",
    title = "{Estimate of the magnetic field strength in heavy-ion collisions}",
    eprint = "0907.1396",
    archivePrefix = "arXiv",
    primaryClass = "nucl-th",
    doi = "10.1142/S0217751X09047570",
    journal = "Int. J. Mod. Phys. A",
    volume = "24",
    pages = "5925--5932",
    year = "2009"
}

@article{Lugones:2015bya,
    author = "Lugones, Germ{\'a}n",
    title = "{From quark drops to quark stars: some aspects of the role of quark matter in compact stars}",
    eprint = "1508.05548",
    archivePrefix = "arXiv",
    primaryClass = "astro-ph.HE",
    doi = "10.1140/epja/i2016-16053-x",
    journal = "Eur. Phys. J. A",
    volume = "52",
    number = "3",
    pages = "53",
    year = "2016"
}

@article{Nambu:1961tp,
    author = "Nambu, Yoichiro and Jona-Lasinio, G.",
    editor = "Eguchi, T.",
    title = "{Dynamical Model of Elementary Particles Based on an Analogy with Superconductivity. 1.}",
    doi = "10.1103/PhysRev.122.345",
    journal = "Phys. Rev.",
    volume = "122",
    pages = "345--358",
    year = "1961"
}

@article{Nambu:1961fr,
    author = "Nambu, Yoichiro and Jona-Lasinio, G.",
    editor = "Eguchi, T.",
    title = "{Dynamical model of elementary particles based on an analogy with superconductivity. II.}",
    doi = "10.1103/PhysRev.124.246",
    journal = "Phys. Rev.",
    volume = "124",
    pages = "246--254",
    year = "1961"
}

@article{Vogl:1991qt,
    author = "Vogl, U. and Weise, W.",
    title = "{The Nambu and Jona Lasinio model: Its implications for hadrons and nuclei}",
    reportNumber = "TPR-91-6",
    doi = "10.1016/0146-6410(91)90005-9",
    journal = "Prog. Part. Nucl. Phys.",
    volume = "27",
    pages = "195--272",
    year = "1991"
}

@article{Klevansky:1992qe,
    author = "Klevansky, S. P.",
    title = "{The Nambu-Jona-Lasinio model of quantum chromodynamics}",
    doi = "10.1103/RevModPhys.64.649",
    journal = "Rev. Mod. Phys.",
    volume = "64",
    pages = "649--708",
    year = "1992"
}

@article{Hatsuda:1994pi,
    author = "Hatsuda, Tetsuo and Kunihiro, Teiji",
    title = "{QCD phenomenology based on a chiral effective Lagrangian}",
    eprint = "hep-ph/9401310",
    archivePrefix = "arXiv",
    reportNumber = "UTHEP-270, RYUTHP-94-1",
    doi = "10.1016/0370-1573(94)90022-1",
    journal = "Phys. Rept.",
    volume = "247",
    pages = "221--367",
    year = "1994"
}

@article{Dumm:2021vop,
    author = "Dumm, D. Gomez and Carlomagno, J. P. and Scoccola, N. N.",
    title = "{Strong-interaction matter under extreme conditions from chiral quark models with nonlocal separable interactions}",
    eprint = "2101.09574",
    archivePrefix = "arXiv",
    primaryClass = "hep-ph",
    doi = "10.3390/sym13010121",
    journal = "Symmetry",
    volume = "13",
    number = "1",
    pages = "121",
    year = "2021"
}

@article{Fukushima:2003fw,
    author = "Fukushima, Kenji",
    title = "{Chiral effective model with the Polyakov loop}",
    eprint = "hep-ph/0310121",
    archivePrefix = "arXiv",
    reportNumber = "MIT-CTP-3424",
    doi = "10.1016/j.physletb.2004.04.027",
    journal = "Phys. Lett. B",
    volume = "591",
    pages = "277--284",
    year = "2004"
}

@article{Contrera:2007wu,
    author = "Contrera, Gustavo A. and Gomez Dumm, Daniel and Scoccola, Norberto N.",
    title = "{Nonlocal SU(3) chiral quark models at finite temperature: The Role of the Polyakov loop}",
    eprint = "0711.0139",
    archivePrefix = "arXiv",
    primaryClass = "hep-ph",
    doi = "10.1016/j.physletb.2008.01.069",
    journal = "Phys. Lett. B",
    volume = "661",
    pages = "113--117",
    year = "2008"
}

@article{Hell:2008cc,
    author = "Hell, T. and Roessner, Simon and Cristoforetti, M. and Weise, W.",
    title = "{Dynamics and thermodynamics of a non-local PNJL model with running coupling}",
    eprint = "0810.1099",
    archivePrefix = "arXiv",
    primaryClass = "hep-ph",
    doi = "10.1103/PhysRevD.79.014022",
    journal = "Phys. Rev. D",
    volume = "79",
    pages = "014022",
    year = "2009"
}

@article{Carlomagno:2013ona,
    author = "Carlomagno, J. P. and G{\'o}mez Dumm, D. and Scoccola, N. N.",
    title = "{Deconfinement and chiral restoration in nonlocal SU(3) chiral quark models}",
    eprint = "1305.2969",
    archivePrefix = "arXiv",
    primaryClass = "hep-ph",
    doi = "10.1103/PhysRevD.88.074034",
    journal = "Phys. Rev. D",
    volume = "88",
    number = "7",
    pages = "074034",
    year = "2013"
}

@article{Karsch:2003jg,
    author = "Karsch, F. and Laermann, E.",
    editor = "Hwa, Rudolph C. and Wang, Xin-Nian",
    title = "{Thermodynamics and in medium hadron properties from lattice QCD}",
    eprint = "hep-lat/0305025",
    archivePrefix = "arXiv",
    pages = "1--59",
    month = "5",
    year = "2003"
}

@article{Bali:2011qj,
    author = "Bali, G. S. and Bruckmann, F. and Endrodi, G. and Fodor, Z. and Katz, S. D. and Krieg, S. and Schafer, A. and Szabo, K. K.",
    title = "{The QCD phase diagram for external magnetic fields}",
    eprint = "1111.4956",
    archivePrefix = "arXiv",
    primaryClass = "hep-lat",
    doi = "10.1007/JHEP02(2012)044",
    journal = "JHEP",
    volume = "02",
    pages = "044",
    year = "2012"
}

@article{Bali:2012zg,
    author = "Bali, G. S. and Bruckmann, F. and Endrodi, G. and Fodor, Z. and Katz, S. D. and Schafer, A.",
    title = "{QCD quark condensate in external magnetic fields}",
    eprint = "1206.4205",
    archivePrefix = "arXiv",
    primaryClass = "hep-lat",
    doi = "10.1103/PhysRevD.86.071502",
    journal = "Phys. Rev. D",
    volume = "86",
    pages = "071502",
    year = "2012"
}

@article{Farias:2014eca,
    author = "Farias, R. L. S. and Gomes, K. P. and Krein, G. I. and Pinto, M. B.",
    title = "{Importance of asymptotic freedom for the pseudocritical temperature in magnetized quark matter}",
    eprint = "1404.3931",
    archivePrefix = "arXiv",
    primaryClass = "hep-ph",
    doi = "10.1103/PhysRevC.90.025203",
    journal = "Phys. Rev. C",
    volume = "90",
    number = "2",
    pages = "025203",
    year = "2014"
}

@article{Ferreira:2014kpa,
    author = "Ferreira, M. and Costa, P. and Louren{\c{c}}o, O. and Frederico, T. and Provid{\^e}ncia, C.",
    title = "{Inverse magnetic catalysis in the (2+1)-flavor Nambu-Jona-Lasinio and Polyakov-Nambu-Jona-Lasinio models}",
    eprint = "1404.5577",
    archivePrefix = "arXiv",
    primaryClass = "hep-ph",
    doi = "10.1103/PhysRevD.89.116011",
    journal = "Phys. Rev. D",
    volume = "89",
    number = "11",
    pages = "116011",
    year = "2014"
}

@article{Pagura:2016pwr,
    author = "Pagura, V. P. and Gomez Dumm, D. and Noguera, S. and Scoccola, N. N.",
    title = "{Magnetic catalysis and inverse magnetic catalysis in nonlocal chiral quark models}",
    eprint = "1609.02025",
    archivePrefix = "arXiv",
    primaryClass = "hep-ph",
    doi = "10.1103/PhysRevD.95.034013",
    journal = "Phys. Rev. D",
    volume = "95",
    number = "3",
    pages = "034013",
    year = "2017"
}

@article{dumm2017strong,
    author = "G{\'o}mez Dumm, D. and Izzo Villafa{\~n}e, M. F. and Noguera, S. and Pagura, V. P. and Scoccola, N. N.",
    title = "{Strong magnetic fields in nonlocal chiral quark models}",
    eprint = "1709.04742",
    archivePrefix = "arXiv",
    primaryClass = "hep-ph",
    doi = "10.1103/PhysRevD.96.114012",
    journal = "Phys. Rev. D",
    volume = "96",
    number = "11",
    pages = "114012",
    year = "2017"
}

@article{Carlomagno:2023clk,
    author = "Carlomagno, J. P. and Ferraris, S. A. and Gomez Dumm, D. and Grunfeld, A. G.",
    title = "{T-{\ensuremath{\mu}} quark matter phase transitions and critical endpoint in nonlocal PNJL models under a strong magnetic field}",
    eprint = "2305.15540",
    archivePrefix = "arXiv",
    primaryClass = "hep-ph",
    doi = "10.1103/PhysRevD.108.056029",
    journal = "Phys. Rev. D",
    volume = "108",
    number = "5",
    pages = "056029",
    year = "2023"
}

@article{Avancini:2012ee,
    author = "Avancini, Sidney S. and Menezes, Debora P. and Pinto, Marcus B. and Providencia, Constanca",
    title = "{The QCD Critical End Point Under Strong Magnetic Fields}",
    eprint = "1202.5641",
    archivePrefix = "arXiv",
    primaryClass = "hep-ph",
    doi = "10.1103/PhysRevD.85.091901",
    journal = "Phys. Rev. D",
    volume = "85",
    pages = "091901",
    year = "2012"
}

@article{Ferrari:2012yw,
    author = "Ferrari, Gabriel N. and Garcia, Andre F. and Pinto, Marcus B.",
    title = "{Chiral Transition Within Effective Quark Models Under Magnetic Fields}",
    eprint = "1207.3714",
    archivePrefix = "arXiv",
    primaryClass = "hep-ph",
    doi = "10.1103/PhysRevD.86.096005",
    journal = "Phys. Rev. D",
    volume = "86",
    pages = "096005",
    year = "2012"
}

@article{Costa:2013zca,
    author = "Costa, Pedro and Ferreira, M{\'a}rcio and Hansen, Hubert and Menezes, D{\'e}bora P. and Provid{\^e}ncia, Constan{\c{c}}a",
    title = "{Phase transition and critical end point driven by an external magnetic field in asymmetric quark matter}",
    eprint = "1307.7894",
    archivePrefix = "arXiv",
    primaryClass = "hep-ph",
    doi = "10.1103/PhysRevD.89.056013",
    journal = "Phys. Rev. D",
    volume = "89",
    number = "5",
    pages = "056013",
    year = "2014"
}

@article{Costa:2015bza,
    author = "Costa, Pedro and Ferreira, M{\'a}rcio and Menezes, D{\'e}bora P. and Moreira, Jo{\~a}o and Provid{\^e}ncia, Constan{\c{c}}a",
    title = "{Influence of the inverse magnetic catalysis and the vector interaction in the location of the critical end point}",
    eprint = "1508.07870",
    archivePrefix = "arXiv",
    primaryClass = "hep-ph",
    doi = "10.1103/PhysRevD.92.036012",
    journal = "Phys. Rev. D",
    volume = "92",
    number = "3",
    pages = "036012",
    year = "2015"
}

@article{Grunfeld:2017dfu,
    author = "Grunfeld, A. G. and Lugones, G.",
    title = "{Finite size effects in strongly interacting matter at zero chemical potential from Polyakov loop Nambu-Jona-Lasinio model in the light of lattice data}",
    eprint = "1711.07559",
    archivePrefix = "arXiv",
    primaryClass = "hep-ph",
    doi = "10.1140/epjc/s10052-018-6113-5",
    journal = "Eur. Phys. J. C",
    volume = "78",
    number = "8",
    pages = "640",
    year = "2018"
}

@article{Grunfeld:2024ihq,
    author = "Grunfeld, Ana Gabriela and Izzo Villafa{\~n}e, Mar{\'\i}a Florencia and Lugones, Germ{\'a}n",
    title = "{Surface and Curvature Tensions of Cold, Dense Quark Matter: A Term-by-Term Analysis Within the Nambu{\textendash}Jona{\textendash}Lasinio Model}",
    eprint = "2407.05606",
    archivePrefix = "arXiv",
    primaryClass = "nucl-th",
    doi = "10.3390/universe11020029",
    journal = "Universe",
    volume = "11",
    number = "2",
    pages = "29",
    year = "2025"
}

@article{Ferraris:2021vun,
    author = "Ferraris, S. A. and Dumm, D. Gomez and Grunfeld, A. G. and Scoccola, N. N.",
    title = "{Cold magnetized quark matter at finite density in a nonlocal chiral quark model}",
    eprint = "2103.00982",
    archivePrefix = "arXiv",
    primaryClass = "hep-ph",
    doi = "10.1140/epja/s10050-021-00463-2",
    journal = "Eur. Phys. J. A",
    volume = "57",
    number = "4",
    pages = "141",
    year = "2021"
}

@article{Ratti:2005jh,
    author = "Ratti, Claudia and Thaler, Michael A. and Weise, Wolfram",
    title = "{Phases of QCD: Lattice thermodynamics and a field theoretical model}",
    eprint = "hep-ph/0506234",
    archivePrefix = "arXiv",
    doi = "10.1103/PhysRevD.73.014019",
    journal = "Phys. Rev. D",
    volume = "73",
    pages = "014019",
    year = "2006"
}

@article{Schaefer:2007pw,
    author = "Schaefer, Bernd-Jochen and Pawlowski, Jan M. and Wambach, Jochen",
    title = "{The Phase Structure of the Polyakov--Quark-Meson Model}",
    eprint = "0704.3234",
    archivePrefix = "arXiv",
    primaryClass = "hep-ph",
    doi = "10.1103/PhysRevD.76.074023",
    journal = "Phys. Rev. D",
    volume = "76",
    pages = "074023",
    year = "2007"
}

@article{Sasaki:2012bi,
    author = "Sasaki, Chihiro and Redlich, Krzysztof",
    title = "{An Effective gluon potential and hybrid approach to Yang-Mills thermodynamics}",
    eprint = "1204.4330",
    archivePrefix = "arXiv",
    primaryClass = "hep-ph",
    doi = "10.1103/PhysRevD.86.014007",
    journal = "Phys. Rev. D",
    volume = "86",
    pages = "014007",
    year = "2012"
}

@article{Ferreira:2017wtx,
    author = "Ferreira, M{\'a}rcio and Costa, Pedro and Provid{\^e}ncia, Constan{\c{c}}a",
    title = "{Multiple critical endpoints in magnetized three flavor quark matter}",
    eprint = "1712.08378",
    archivePrefix = "arXiv",
    primaryClass = "hep-ph",
    doi = "10.1103/PhysRevD.97.014014",
    journal = "Phys. Rev. D",
    volume = "97",
    number = "1",
    pages = "014014",
    year = "2018"
}

@article{Menezes:2008qt,
    author = "Menezes, D. P. and Benghi Pinto, M. and Avancini, S. S. and Perez Martinez, A. and Providencia, C.",
    title = "{Quark matter under strong magnetic fields in the Nambu-Jona-Lasinio Model}",
    eprint = "0811.3361",
    archivePrefix = "arXiv",
    primaryClass = "nucl-th",
    doi = "10.1103/PhysRevC.79.035807",
    journal = "Phys. Rev. C",
    volume = "79",
    pages = "035807",
    year = "2009"
}
\end{adjustwidth}
\end{document}